\documentclass[letterpaper]{article} 

\usepackage[usenames,dvipsnames]{color}
\usepackage{bm}
\usepackage[title]{appendix}
\usepackage{graphicx,fancyhdr,lscape,dsfont,amsmath,amssymb} 
\usepackage{color, amsthm}
\usepackage{longtable}
\usepackage{psfrag}
\usepackage{subfigure}
\usepackage{tabularx}
\usepackage{boldline,multirow}
\usepackage{stmaryrd}
\usepackage[sl]{caption}
\usepackage{appendix}
\usepackage{soul}
\usepackage{cancel}
\usepackage{ulem} 
\usepackage{sidecap}
\sidecaptionvpos{figure}{c}

\allowdisplaybreaks

\newcommand{\vect}[1]{\vec{#1}} 
\newcommand{\tensor}[1]{  {\bm {#1}} } 

\newcommand{\trace}[1]{ {\rm tr} \left[ \, {#1} \, \right] }

\newcommand{\divergence}[1]{ {\rm div} \left[ \, {#1} \, \right] }
\newcommand{\gradient}[1]{ {\rm \nabla} \left[ \, {#1} \, \right] }

\newcommand{\diffusivity}{\mbox{${\rm D} \mskip-8mu  | \,$}}

\renewcommand{\em}[1]{\it{#1}}

\begin{document}

\title{A coupled model of diffusional creep of polycrystalline solids based on climb of dislocations at grain boundaries} 
\author{M. Magri$^{1}$, G. Lemoine$^{2}$, L. Adam$^{2}$ and J. Segurado$^{1,3}$
\\
\\
\begin{small}
$^{1}$ Fundaci\'on IMDEA Materiales, C/ Eric Kandel 2, 
28906, Getafe, Madrid, Spain
\end{small}
 \\
 \\
\begin{small}
$^{2}$ e-Xstream Engineering, Axis Park-Building H,
Rue Emile Francqui 9, B-1435 Mont-Saint-Guibert, Belgium
 \end{small}
 \\
 \\
\begin{small}
$^{3}$ Department of Materials Science, Technical University of Madrid, E.T.S. de Ingenieros de Caminos,
28040, Madrid, Spain
\end{small}
}

\maketitle

\begin{abstract}
A continuum theory based on thermodynamics has been developed for modeling diffusional creep of polycrystalline solids. It consists of a coupled problem of vacancy diffusion and mechanics where the vacancy generation/absorption at grain boundaries is driven by grain boundary dislocations climb. The model is stated in terms of general balance laws and completed by the choice of constitutive equations consistent with classical non-equilibrium thermodynamics. The kinetics of diffusional creep is derived from physically-based mechanisms of climb of dislocations at grain boundaries, thus introducing a dependence of diffusional creep on the density and mobility of boundary dislocations. Several representative examples have been solved using the finite element method and assuming representative volume elements made up of an array of regular-shaped crystals. The effect of stress, temperature, grain size, and grain boundary dislocation mobility is analyzed and compared with classical theories of diffusional creep. The simulation results demonstrate the ability of the present model to reproduce the macroscopic stress and grain size dependence observed under both diffusion and interface controlled regimes as well the evolution of this dependency with the temperature. In addition, the numerical implementation of the model allows to predict the evolution of microscopic fields through the microstructure.


\end{abstract}

\section{Introduction}

At elevated homologous temperatures, the plastic deformation of crystalline solids can be dominated by diffusion-controlled processes since the mobility of atoms and vacancies increases rapidly with temperature. Such mechanisms include the so-called \textit{diffusional creep} in which a plastic deformation results as a consequence of self-diffusion of atoms in individual crystals. Diffusional creep dominates the creep material response at low stress levels compared with the stress required for dislocation glide. 
Moreover, diffusional creep is strongly related to the presence of grain boundaries since grain boundary dislocation climbing is enhanced at these temperatures and this process involves the  generation or absorption of vacancies. The amount of grain boundaries, and therefore the grain size, strongly influences the creep behavior, leading to a grain size dependence of the material of the type ``smaller is weaker" \cite{ARZT19985611}. This size effect, found at high temperature, is opposite to the characteristic grain size effect of polycrystals at room temperature, when deformation is based on dislocation glide and a grain size dependence appears due to dislocation pilling up \cite{HAOUALA201872,HAOUALA2019103755}. 

 The possibility of diffusion-induced creep in polycrystalline solids was first proposed in the pioneering works of Nabarro \cite{Nabarro1948}, Herring \cite{Herring1950} and Coble \cite{Coble1963}. In their theories, a steady state creep rate $\dot{\varepsilon}$ was estimated by considering two concurrent processes, namely the operation of vacancy source and sink at grain-boundary (GB) and the diffusion between sources and sinks. If the rate of generation/absorption of vacancies at GB is much faster than the rate of diffusion, the latter is the rate controlling process and boundaries act as perfect sinks and sources for vacancies. In such a case, $\dot{\varepsilon} \propto {\sigma}/{d^2}$ \cite{Herring1950} if the diffusion takes place through the lattice, and $\dot{\varepsilon} \propto {\sigma}/{d^3}$ \cite{Coble1963} in case of dominant GB diffusion ($d$ is the grain size and $\sigma$ the applied stress).
 
 Subsequently, Ashby \cite{Ashby1969,Ashby1972}, Burton \cite{Burton1971}, and Artz et al. \cite{Artz1983} relaxed the hypothesis of perfect boundaries in the calculation of diffusional creep rates. In particular, Ashby \cite{Ashby1969} proposed that the operation of vacancy sources and sinks is the consequence of climb of dislocations along GB. In this view, the rate of vacancy creation and annihilation can be limited either if the GB dislocation density is insufficient \cite{Burton1971}, or if the GB dislocation mobility is affected by the presence of impurities, solutes, or precipitates  \cite{Artz1983}. In both cases, a $\dot{\varepsilon} \propto {\sigma^2}/{d}$ relation was predicted when creep deformation is interface controlled, showing a higher stress dependence than in the case of perfect boundaries. 
 
More recently, numerous studies have proposed continuum models for describing diffusional creep of polycrystalline aggregates. In the seminal work of Needleman and Rice \cite{Needleman1980}, a diffusion-driven plastic strain was incorporated at the interfaces among grains with perfect GBs. The proposed formulation was established in terms of a variational principle and solved numerically through the finite element method. This formulation was extended by Cocks \cite{Cocks1992} in order to account for imperfect GBs. Similarly, the transient response of polycrystalline aggregates separated by sharp GBs, has been analyzed when GB-sliding is coupled with GB-diffusion by Wei et al. \cite{Wei2008}. However, all these studies consider that atomic diffusion takes place only at the interface among grains.

Continuum formulations for the coupled phenomena of diffusion and mechanics in the overall polycrystalline solid can also be found in the literature. Garikipati et al. \cite{Garikipati2001} developed a lattice-based model in which the diffusional process is posed in terms of vacancy diffusion. In this work, a plastic strain was assigned in the GB regions to model the effect of creep and it was introduced as result of atoms diffusing towards the boundary without considering the dislocation-based mechanisms underlying the mechanism. Moreover, the rate of GB plastic strain was introduced through 
a penalty-like parameter in order to retrieve the case of perfect  boundaries, leading to a diffusion controlled process independent on the grain boundary characteristics. A similar approach for modeling diffusional creep was pursued recently by Villani et al. \cite{Villani2015}. The model proposed also included creep deformation in the grain interior due to inhomogeneous flux of vacancies and dislocation plasticity. However, the resulting governing equations have not been derived by following a clear thermodynamical framework, which is crucial when coupled multi-physical processes are considered. In addition, as in  \cite{Garikipati2001}, vacancy generation/annihilation in the grain boundary is assumed instantaneous such that only diffusion controlled regime can be considered.

A remarkable study on the thermodynamics of diffusional creep can be found in the work by Mishin et al. \cite{MIshin2013}, where the creep deformation was assumed to be driven by mechanisms involving site generation/annihilation and flux of vacancies. The rate of  dissipative processes was identified based on thermodynamic restrictions but the kinetics of GB processes was again not addressed in detail. Similarly, a rigorous thermodynamical framework for diffusion and creep of crystalline systems was proposed by Svoboda et al. \cite{Svoboda2006}. However, their theory aimed to model the chemo-mechanical processes involved with dislocation climb in the grain interior, rather than focusing on GB diffusional creep. 

The objective of this paper is to develop a crystal-level continuum model capable of describing diffusional creep of polycrystalline solids based on GB dislocation mechanisms. In particular, the following aspects characterize the present study.

\begin{itemize}

\item
The theory is formulated as a coupled problem of vacancy diffusion and mechanics where vacancy sink and sources in the grain boundaries are linked to dislocation climb. Governing equations are derived using a consistent thermodynamic framework as desirable for this class of multi-physics processes. In such a way, the kinetics of GB processes is correctly identified as function of its conjugated thermodynamic force.

\item
Diffusional creep is assumed to occur by climb of GB dislocations as proposed by Ashby \cite{Ashby1969,Ashby1972}. Differently from most existing continuum models, the creep kinetics is derived from physically-based mechanisms of dislocations at the grain boundary. This allows to study the evolution of diffusional creep as a function of grain boundary dislocation properties, such as their mobility and the density within the GB. The formulation is able to describe creep processes controlled by diffusion or by vacancy generation at the interfaces.

\item
In the scope of the present study, the solid is subjected to conditions such that diffusional creep is the dominant plastic deformation mechanism. Accordingly, other inelastic contributions, such as dislocation plasticity in the lattice and GB sliding, are not considered here.

\item
The impact of grain size, applied stress, temperature, and GB dislocation mobility is studied through a series of representative examples. To this end, the proposed governing equations are solved numerically through the finite element method. 

\end{itemize}

The paper is organized as follows. The fundamental hypothesis of the model along with the basic conservation laws and equations describing the mechanics of diffusional creep will be discussed in Section \ref{sec:balance_laws}. Thermodynamic restrictions are derived in Section \ref{sec:thermodynamics} from thermodynamic principles stated in terms of conservation of energy and entropy imbalance. Subsequently, in Section \ref{sec:constitutive_theory}, constitutive theory provides consistent specifications for diffusion, mechanical stress, and kinetics of GB processes. The final form of governing equations is summarized in Section \ref{sec:governing_equations}, 
where the extreme limits of diffusion controlled and reaction controlled creep are identified. Section \ref{sec:numerical_ex} is devoted to the discussion of numerical examples, while Section \ref{sec:conclusions} closes the paper with some concluding remarks. 

\section{Conservation laws and diffusional creep} \label{sec:balance_laws}
A crystalline solid is modeled as a binary system consisting of two diffusing species, namely atoms $(A)$ and vacancies $(V)$. Atoms and vacancies reside in specific sites called lattice sites (L). For simplicity, pure crystals or solid solutions with small content of second species are considered, so only one type of atom need to be considered for the thermodynamics. Atoms are conserved in the solid while vacancies are not because of vacancy emission/absorption at climbing dislocations in GBs. 



\begin{figure}[htbp]
\centering
 \includegraphics[width=10cm]{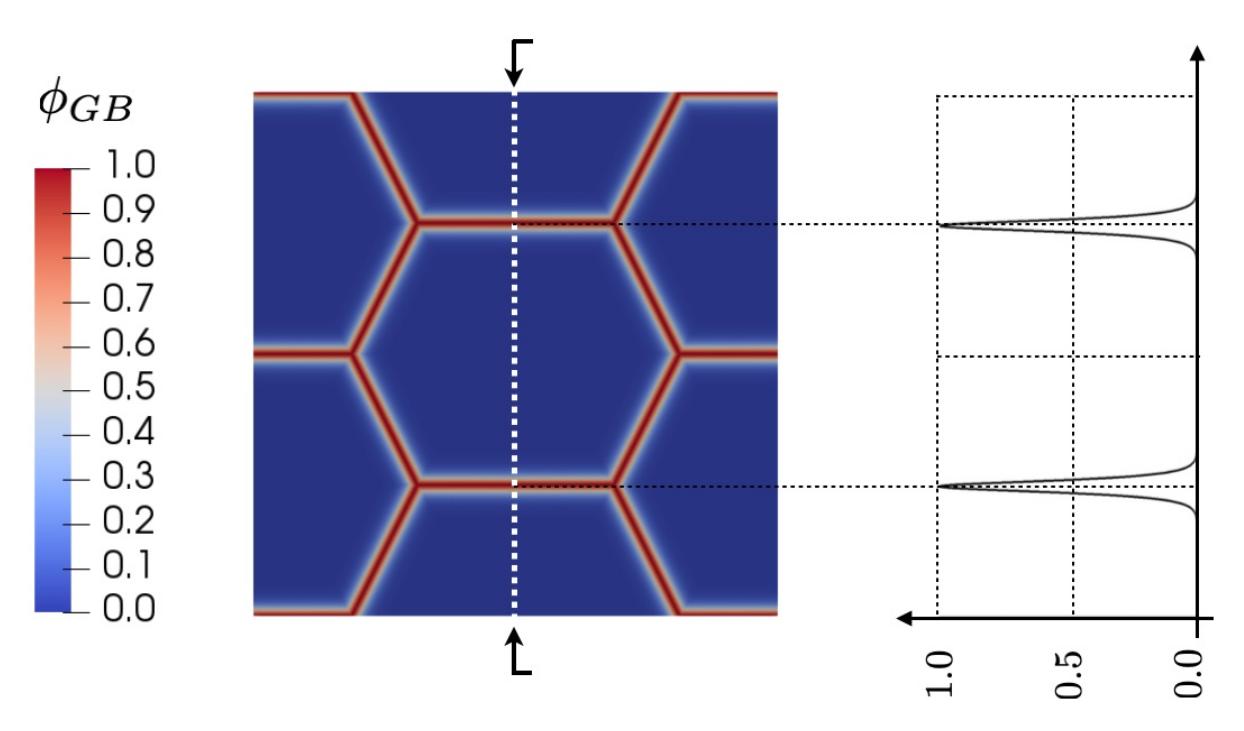}
\caption{\em{Plot of the phase-field function defined in Eq. (\ref{eq:GB_indicator_fun}) for an idealized polycrystalline aggregate. By choosing $r_G=5.3$, it results that $\phi_{GB} (d_{GB}/2) = \phi_{GB} (-d_{GB}/2) = 0.01$.}}
\label{fig:GBfunction}
\end{figure}

Following \cite{Villani2015}, a diffuse description of grain boundaries is adopted here by introducing the following phase-field function 

\begin{equation} \label{eq:GB_indicator_fun}
\phi_{GB} \left( \bar{d} \right) = \left[ \text{cosh} \left( \frac{r_G \, 2 \bar{d}}{d_{GB}}  \right) \right]^{-1} \, ,
\end{equation}

\noindent
where $\bar{d}$ is the distance of any point in the lattice to the closest grain boundary, $d_{GB}$ is the thickness of the grain diffuse boundary, and $r_G$ is a coefficient. As GB migration is not considered in this study, the phase-field function is time-independent. According to Eq. \eqref{eq:GB_indicator_fun}, $\phi_{GB}$ is a smooth function of the distance to the closest GB with values ranging from 0 to 1. For a point situated in a grain boundary $\phi_{GB}=1$, while for points in the lattice $\phi_{GB} \rightarrow 0$. In the diffuse interface description, GBs possess a narrow region of finite thickness, whose magnitude depends on $d_{GB}$ and $r_{GB}$. An example of the considered function $\phi_{GB}$ for an idealized polycrystal is reported in Fig.  \ref{fig:GBfunction}.

\subsection{Balance of diffusing species}

\medskip

Based on the previous hypothesis, the conservation laws for atoms and vacancies read

\begin{subequations} \label{eq:mass_balance}
\begin{align}
& \dot{c}_A + \text{div} \left[ \vec{h}_A \right] = 0 \, , \\
\nonumber \\
 & \dot{c}_V+ \text{div} \left[ \vec{h}_V \right] = \phi_{GB}  \, s_V \, .
\end{align}
\end{subequations}

\noindent
In equations \eqref{eq:mass_balance}, $c_A$ and $c_V$ are molar concentrations - i.e. the number of moles per unit volume - of atoms and vacancies, respectively; a superposed dot indicates a partial time derivative; $\vec{h}_A$ and $\vec{h}_V$ are molar fluxes; $s_V$ is the molar rate of generation/absorption of vacancies at grain boundaries. Since atoms diffuse because of vacancy diffusion, the following condition holds

\begin{equation} \label{eq:fluxes}
\vec{h}_A = - \vec{h}_V \, .
\end{equation}

\noindent 
In addition, all the available lattice sites are occupied either by atoms or vacancy, thus

\begin{equation} \label{eq:lattice_constraint}
c_A + c_V = c_L \, ,
\end{equation}

\noindent 
where $c_L$ is the molar concentration of lattice sites. By combining equations \eqref{eq:mass_balance}, \eqref{eq:fluxes}, and \eqref{eq:lattice_constraint} it can be easily proved that

\begin{equation} \label{eq:lattice_rate}
\dot{c}_L = \phi_{GB}  \, s_V \, .
\end{equation}

\noindent
It thus results that lattice sites can be altered at GB because of vacancies creation or annihilation. 

\subsection{Balance of momentum}

Assuming the inertial forces to be negligible, the balance of momentum yields

\begin{equation} \label{eq:balance_of _momentum}
\text{div} \left[ \boldsymbol{\sigma} \right] + \vec{b} = \vec{0} \, ,
\end{equation}
 
\noindent
where $\boldsymbol{\sigma}$ is the symmetric Cauchy stress tensor, while $\vec{b}$ is the body force per unit volume.

\subsection{Mechanics of diffusional creep} \label{subsec:mechanics_of_creep}

\begin{figure}[htbp]
\centering
 \includegraphics[width=16cm]{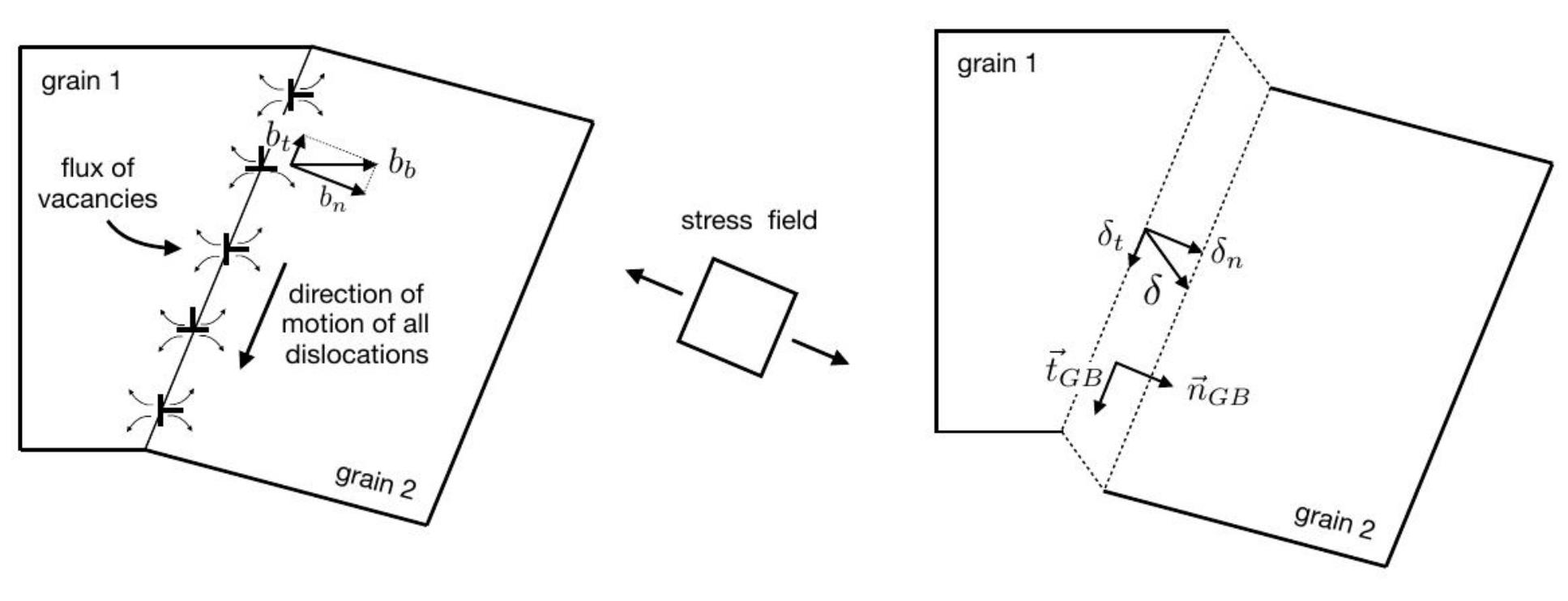}
\caption{\em{(a) Schematic of the boundary mechanisms of diffusional creep involving the motion of dislocations along grain boundaries \cite{Ashby1972}. (b) Relative translation of the grains that meet at the boundary due to a general orientation of GB dislocations.}}
\label{fig:GBclimb}
\end{figure}

The microscopic aspects of diffusional creep are related to the motion of boundary defects as suggested by Ashby \cite{Ashby1972}. The key idea is that the operation of vacancies emission/absorption at GBs originates from climbing boundary dislocations, which are actually the sinks and sources. GB dislocations have Burgers' vectors ($b_b$) which are not, usually, lattice vectors, and therefore their motion is constrained to take place in the boundary plane. The application of a stress field normal to the grain boundary makes dislocations move along the boundary plane, as reported in Fig. \ref{fig:GBclimb}a. If so, dislocations move by a combination of glide and climb motions, which depends on the orientation of the Burgers' vector with respect to the GB. Only boundary dislocations with a component of their Burgers' vector normal to the boundary ($b_n$) move non-conservatively, thus emitting/absorbing vacancies. This causes a flux of vacancies and a counter-flux of atoms. 

At a continuum scale, the motion of boundary dislocations involves, in general, both relative normal and shear translation of the crystals that meet at the boundary (see Fig. \ref{fig:GBclimb}b). Their relative magnitude at any point depends on the orientation of the Burgers' vector relative to the boundary. In particular, the amount of normal displacement is the non-conservative part that originates from climb of boundary dislocations. For the scope of this paper, attention is paid to the normal translation only, being the latter the component involved with emission/absorption of vacancies, and thus relevant for diffusional creep. Therefore, we assume that boundary dislocations move only by climb along the grain boundary. A similar idealization of the boundary structure is usually adopted while considering the discrete source and sink model of diffusional creep, as for example in \cite{Artz1983}. Based on the considered assumptions, the plastic deformation rate due to diffusional creep at GB is defined as

\begin{equation} \label{eq:def_climb_deformation}
\dot{\boldsymbol{\varepsilon}}^{diff} =  \, \dot{\beta} \;  \left( \vec{n}_{GB} \otimes \vec{n}_{GB} \right)  \, ,
\end{equation}

\noindent
where $\vec{n}_{GB}$ is the unit normal of the grain boundary, while $\dot{\beta}$ denotes the rate of diffusional creep.

The strain rate due to diffusional creep $\dot{\boldsymbol{\varepsilon}}^{diff}$ is uniaxial and possesses both deviatoric and volumetric components. The latter one can be easily calculated as follows

\begin{equation} \nonumber
\text{tr} [ \dot{\boldsymbol{\varepsilon}}^{diff} ] =  \dot{\beta} \, . 
\end{equation}

\noindent
Therefore, according to \cite{HirthLothe1967}, the rate of diffusional creep $\dot{\beta}$ can be related to the rate of emission/absorption of vacancy as

\begin{equation} \label{eq:relationship_climb_source}
  s_V  = \frac{ \dot{\beta} }{ v_A } \, ,
\end{equation}

\noindent 
being $v_A$ the molar volume, i.e. the volume of a mole of atoms. In addition, by exploiting Eq. \eqref{eq:lattice_rate}, it results that

\begin{equation} \label{eq:relationship_climb_lattice}
   \dot{c}_L  = \phi_{GB} \, \frac{ \dot{\beta} }{ v_A } \, .
\end{equation}

\noindent
Equations \eqref{eq:relationship_climb_source} and \eqref{eq:relationship_climb_lattice}  establish a direct coupling between GB deformation and diffusion of vacancies. 

The accumulation or loss of intrinsic point defects, such as vacancies, also causes a local distortion of the lattice both at grain boundary and in the grain interior. Such deformation is usually considered to be purely volumetric, i.e. the removal of atoms causes a local shrinkage of the lattice. The following eigenstrain $\dot{\boldsymbol{\varepsilon}}^{V}$ is then introduced  

\begin{equation} \nonumber
\dot{\boldsymbol{\varepsilon}}^{V} =  \omega_V \, \dot{c}_V  \boldsymbol{I} \, ,
\end{equation}

\noindent
where $\omega_V$ is the coefficient of chemical expansion of vacancies, i.e. one third of the relaxed volume per mole of vacancies.
To account for the mechanics of diffusional creep, the strain tensor $\boldsymbol{\varepsilon}$ is decomposed additively in

\begin{equation} 
\boldsymbol{\varepsilon} = \boldsymbol{\varepsilon}^{el} + \phi_{GB} \, \boldsymbol{\varepsilon}^{diff} + \boldsymbol{\varepsilon}^{V} \, , 
\end{equation}

\noindent
where $\boldsymbol{\varepsilon}^{el}$ refers to the elastic part of the strain.
The GB phase-field function $\phi_{GB}$ has been introduced in order to restrict the impact of $\boldsymbol{\varepsilon}^{diff}$ to GBs only. Note that additional inelastic mechanisms, such as dislocations plasticity in the lattice and GB sliding are not considered in this study. 

\section{Thermodynamics} \label{sec:thermodynamics}
In this section, the classical laws of thermodynamics will be stated for the model at hand. Notation and basic assumptions follow from the paper by Salvadori et al. \cite{SalvadoriEtAlJBOH2016}, to which the reader can refer for further details. 
In what follows, ${\cal{P}}$ indicates a generic subregion of the solid with closed boundary ${\partial \cal{P}}$. Note that ${\cal{P}}$ might include bulk crystals and GBs. 

\subsection{Energy balance}

The energy balance for the problem at hand, for quasi-static interactions, yields

\begin{equation} \label{eq:energy_balance}
\dot{\cal{U}} ( {\cal{P}} ) = {\cal{W}}_u ({\cal{P}}) + {\cal{Q}}_u ({\cal{P}}) + {\cal{T}}_u ({\cal{P}}) \, ,
\end{equation}

\noindent
with ${\cal{U}}$ denoting the net internal energy of $\cal{P}$,  ${\cal{W}}_u$ the mechanical external power, $ {\cal{Q}}_u$ the power due to heat transfer, and ${\cal{T}}_u$ the power due to mass transfer. The individual contributions read

\begin{subequations} 
\begin{align*}
&{\cal{U}} ( {\cal{P}} ) = \int_{\cal{P}} u \, {\text{d}} V \, , \\
 \nonumber \\
 &{\cal{W}}_u ( {\cal{P}} ) = \int_{\cal{P}} \vec{b} \cdot \vec{v} \, {\text{d}} V + \int_{\partial \cal{P}} \vec{t} \cdot \vec{v} \, {\text{d}} A , \\
 \nonumber \\
 &{\cal{Q}}_u ( {\cal{P}} ) = \int_{\cal{P}} s_q  \, {\text{d}} V -  \int_{\partial \cal{P}} \vec{q} \cdot \vec{n} \, {\text{d}} A , \\
 \nonumber \\
 &{\cal{T}}_u ( {\cal{P}} ) =  - \int_{\partial \cal{P}} {}^u\mu_A \, \vec{h}_A  \cdot \vec{n} \, {\text{d}} A -  \int_{\partial \cal{P}} {}^u\mu_V \, \vec{h}_V \cdot \vec{n} \, {\text{d}} A , 
\end{align*}
\end{subequations}

\noindent 
where $u$ is the specific internal energy per unit volume, $\vec{t}$ is the surface traction, $\vec{v}$ is the velocity, $s_q$ is the rate of energy per unit volume at which heat is generated by sources, $\vec{q}$ is the heat flux, and $\vec{n}$ the outward normal of ${\partial \cal{P}}$. In the mass contribution ${\cal{T}}_u$, scalars ${}^u\mu_A$ and ${}^u\mu_V$ denote the change in specific energy provided by a unit supply of moles of atoms and vacancies, respectively. 

Standard application of the divergence theorem, balance laws \eqref{eq:mass_balance}, \eqref{eq:balance_of _momentum}, and Eq. \eqref{eq:fluxes},  leads from \eqref{eq:energy_balance} to

\begin{equation} \label{eq:energy_balance2}
\int_{\cal{P}} \dot{u} \, {\text{d}} V  = \int_{\cal{P}} \boldsymbol{\sigma} : \dot{\boldsymbol{\varepsilon}} \, {\text{d}} V +  \int_{\cal{P}} s_q - \text{div} \left[ \vec{q} \right]   \, {\text{d}} V + \int_{\cal{P}} \left( {}^u\mu_V - {}^u\mu_A \right) \left( \dot{c}_V - \phi_{GB} s_V \right)  - \nabla \left[ {}^u\mu_V - {}^u\mu_A \right] \cdot \vec{h}_V \, {\text{d}} V  .
\end{equation}

\noindent
where $\boldsymbol{\varepsilon}$ is the strain tensor, i.e. $\dot{\boldsymbol{\varepsilon}} = \text{sym} \left[ \nabla \left[ \vec{v} \right] \right]$. Equation \eqref{eq:energy_balance2} must hold for any region $\cal{P}$, since the latter is arbitrary. The local form of the energy balance is then stated as follows

\begin{equation} \label{eq:local_energy_balance}
\dot{u}  =\boldsymbol{\sigma} : \dot{\boldsymbol{\varepsilon}} + s_q - \text{div} \left[ \, \vec{q} \, \right]   +  \left( {}^u\mu_V - {}^u\mu_A \right) \left( \dot{c}_V - \phi_{GB} s_V \right)  - \nabla \left[ {}^u\mu_V - {}^u\mu_A \right] \cdot \vec{h}_V .
\end{equation}

\subsection{Entropy imbalance}

The entropy imbalance for the problem at hand, for quasi-static interactions, yields

\begin{equation} \label{eq:entropy_imbalance}
\dot{\cal{S}} ( {\cal{P}} )  \geq  {\cal{Q}}_\eta ({\cal{P}}) + {\cal{T}}_\eta ({\cal{P}}) \, ,
\end{equation}

\noindent
where ${\cal{S}}$ is the net internal entropy of ${\cal{P}}$, $ {\cal{Q}}_\eta$ is the exchanged entropy per unit time due to heat transfer, and ${\cal{T}}_\eta$ the exchanged entropy per unit time due to mass transfer. The individual contributions read

\begin{subequations} 
\begin{align*}
&{\cal{S}} ( {\cal{P}} ) = \int_{\cal{P}} \eta \, {\text{d}} V \, , \\
 \nonumber \\
 &{\cal{Q}}_\eta ( {\cal{P}} ) = \int_{\cal{P}} \frac{s_q}{T}  \, {\text{d}} V - \int_{\partial \cal{P}} \frac{\vec{q}}{T} \cdot \vec{n} \, {\text{d}} A , \\
 \nonumber \\
 &{\cal{T}}_\eta ( {\cal{P}} ) =  - \int_{\partial \cal{P}} {}^\eta\mu_A \, \vec{h}_A  \cdot \vec{n} \, {\text{d}} A -  \int_{\partial \cal{P}} {}^\eta\mu_V \, \vec{h}_V \cdot \vec{n} \, {\text{d}} A , 
\end{align*}
\end{subequations}

\noindent 
where $\eta$ is the specific internal entropy per unit volume, while $T$ is the absolute temperature. In the mass contribution ${\cal{T}}_\eta$, scalars ${}^\eta\mu_A$ and ${}^\eta\mu_V$ denote the change in specific entropy provided by a unit supply of moles of atoms and vacancies, respectively. 

Standard application of the divergence theorem, balance laws \eqref{eq:mass_balance}, \eqref{eq:balance_of _momentum}, and condition \eqref{eq:fluxes}, leads Eq. \eqref{eq:entropy_imbalance} to its local counterpart

\begin{equation*} 
\dot{\eta}  -  \frac{s_q}{T} + \text{div} \left[  \frac{\vec{q}}{T} \right]  - \left( {}^\eta\mu_V - {}^\eta\mu_A \right) \left( \dot{c}_V - \phi_{GB} s_V \right)  + \nabla \left[ {}^\eta\mu_V - {}^\eta\mu_A \right] \cdot \vec{h}_V \geq 0 \, .
\end{equation*}

\noindent
Taking advantage of identity \eqref{eq:local_energy_balance} and of the sign definiteness of temperature, the local form of the entropy imbalance can be rewritten as follows

\begin{equation} \label{eq:local_entropy_imbalance}
T \dot{\eta} - \dot{u} + \boldsymbol{\sigma} : \dot{\boldsymbol{\varepsilon}} + \mu  \left( \dot{c}_V - \phi_{GB} s_V \right)  - \nabla \left[ \,  \mu  \, \right] \cdot \vec{h}_V  - \frac{1}{T} \vec{q} \cdot \nabla \left[ T \right] -  {}^\eta\mu \, \vec{h}_V \cdot \nabla \left[ T \right] \geq 0 \,,
\end{equation}

\noindent 
where ${}^\eta\mu = {}^\eta \mu_V  - {}^\eta \mu_A $. In eq. \eqref{eq:local_entropy_imbalance}, $\mu = \mu_V - \mu_A $  is the so-called \textit{diffusional potential}, i.e. the difference between the chemical potential of vacancies and the chemical potential of atoms. According to \cite{SalvadoriEtAlJBOH2016}

\begin{gather*}
\mu_A = {}^u\mu_A  -  T \, {}^\eta \mu_A  \quad \text{and} \quad \mu_V = {}^u\mu_V  -  T \, {}^\eta \mu_V . 
\end{gather*} 

%
%

\subsection{Helmholtz free energy and thermodynamic restrictions}

The Helmholtz free energy density per unit volume $\psi$, defined as

\begin{equation*}
\psi = u - T  \, \eta \, ,
\end{equation*}
 
\noindent
will be used henceforth as thermodynamic potential for the present theory. It thus follows that 

\begin{equation*}
\dot{\psi} = \dot{u} - \dot{T} \, \eta - T \, \dot{\eta} \, ,
\end{equation*}

\noindent
which can be inserted in \eqref{eq:local_entropy_imbalance} to obtain the following free energy imbalance

\begin{equation*} 
 \dot{\psi} + \dot{T} \, \eta -  \boldsymbol{\sigma} : \dot{\boldsymbol{\varepsilon}} - \mu  \left( \dot{c}_V - \phi_{GB} s_V \right)  + \nabla \left[\, \mu \,\right] \cdot \vec{h}_V  +  \frac{1}{T} \, \vec{q} \cdot \nabla \left[ T \right] +  {}^\eta\mu \, \vec{h}_V \cdot \nabla \left[ T \right] \leq 0 \, .
\end{equation*}

\noindent
Assuming \textit{isothermal conditions} the final form of the entropy imbalance reads

\begin{equation} \label{eq:free_energy_imbalance}
 \dot{\psi}  -  \boldsymbol{\sigma} : \dot{\boldsymbol{\varepsilon}} - \mu  \left( \dot{c}_V - \phi_{GB} s_V \right)  + \nabla \left[ \, \mu \, \right] \cdot \vec{h}_V   \leq 0 \, .
\end{equation}

In order to model different coupled responses at grain boundaries and in grain interiors, $\psi$ is split into a GB energy $\psi_{GB}$ and a bulk contribution $\psi_{bulk}$ as 

\begin{equation} \label{eq:spli_free_energy}
\psi = ( 1 - \phi_{GB}) \,  \psi_{bulk} +  \phi_{GB} \, \psi_{GB} \, .
\end{equation}

\noindent
We further consider the following functional dependence of the free energies 

\begin{equation} \label{eq:state_variables}
\psi_{bulk} =  \psi_{bulk} \left( c_V \, , \boldsymbol{\varepsilon}^{ce} \right) \qquad  \text{and}  \qquad \psi_{GB} =  \psi_{GB} \left( c_V \, , \boldsymbol{\varepsilon}^{ce} \, , \beta \right) \, ,
\end{equation}

\noindent
where $\boldsymbol{\varepsilon}^{ce} = \boldsymbol{\varepsilon}^{el} + \boldsymbol{\varepsilon}^{V}$ defines the chemo-elastic strain\footnote{This is not the only possible choice, as discussed in \cite{SalvadoriEtAlJBOH2016}. Focusing on the functional dependence on the strain only, the Helmholtz free energy could be written as a function of the whole strain tensor and its inelastic counterpart 
\begin{equation}
\psi = \psi \left( {\boldsymbol{\varepsilon}}, {\boldsymbol{\varepsilon}^{diff}}, ... \right)
\end{equation}}. The internal variable $\beta$, defined in Section \ref{subsec:mechanics_of_creep} accounts for combined chemo-mechanical processes associated with diffusional creep.  Hence, owing to equations \eqref{eq:relationship_climb_source} and \eqref{eq:relationship_climb_lattice}, the evolution of $s_V$ and $c_L$ at GBs can be expressed in terms of $\beta$. Similarly, $c_A$ is fully determined by $c_V $ and $\beta$, thus it will not be considered as an independent field from now on. 
 
 \medskip

Applying a standard chain-rule, the partial time derivative of $\psi$ from \eqref{eq:spli_free_energy} and  \eqref{eq:state_variables} reads

\begin{equation} \label{eq:rate_helmholtz}
\dot{\psi} = (1- \phi_{GB}) \left[  \frac{ \partial \psi_{bulk}}{ \partial c_V} \, \dot{c}_V + \frac{\partial \psi_{bulk}}{ \partial \boldsymbol{\varepsilon}^{ce}} : \dot{\boldsymbol{\varepsilon}}^{ce} \right] + \phi_{GB}  \left[ \frac{ \partial \psi_{GB}}{ \partial c_V} \, \dot{c}_V + \frac{\partial \psi_{GB}}{ \partial \boldsymbol{\varepsilon}^{ce}} : \dot{\boldsymbol{\varepsilon}}^{ce} +  \frac{\partial \psi_{GB}}{ \partial \beta} \, \dot{\beta} \,  \right] \,\, ,
\end{equation}

\noindent 
which, substituted into \eqref{eq:free_energy_imbalance}, gives

\begin{gather*}
\left[ (1- \phi_{GB}) \frac{\partial \psi_{bulk}}{ \partial \boldsymbol{\varepsilon}^{ce}} +  \phi_{GB} \frac{\partial \psi_{GB}}{ \partial \boldsymbol{\varepsilon}^{ce}}  - \boldsymbol{\sigma} \right] : \dot{\boldsymbol{\varepsilon}}^{ce} + 
  \left[ (1- \phi_{GB})  \frac{ \partial \psi_{bulk}}{ \partial c_V} + \phi_{GB} \frac{ \partial \psi_{GB}}{ \partial c_V} - \mu \right] \dot{c}_V \, +  \\
 + \, \phi_{GB}  \left( \frac{\partial \psi_{GB}}{ \partial \beta} + \frac{\mu}{v_A} \right)  \, \dot{\beta} - \phi_{GB} \,\boldsymbol{\sigma} :  \dot{\boldsymbol{\varepsilon}}^{diff}  + \nabla \left[ \mu \right] \cdot \vec{h}_V   \leq 0 \, .
\end{gather*}

\medskip

\noindent
The latter, which is usually referred as the Clausius-Duhem inequality, must hold for any value of the time derivative of $c_V$ and of the chemo-elastic strain ${\boldsymbol{\varepsilon}}^{ce}$. Since they appear linearly in the inequality, the factors multiplying them must be zero. The following restrictions thus apply

\begin{subequations} \label{eq:TD_restrictions1}
\begin{gather}
\boldsymbol{\sigma} = (1- \phi_{GB}) \frac{\partial \psi_{bulk}}{ \partial \boldsymbol{\varepsilon}^{ce}} +  \phi_{GB} \frac{\partial \psi_{GB}}{ \partial \boldsymbol{\varepsilon}^{ce}} \, ,  \\ 
 \nonumber \\ 
 \mu = (1- \phi_{GB})  \frac{ \partial \psi_{bulk}}{ \partial c_V} + \phi_{GB} \frac{ \partial \psi_{GB}}{ \partial c_V} \, .  \label{eq:TD_restrictions_mu}
\end{gather}
\end{subequations}

\noindent
The remaining terms constitute the total energetic dissipation

\begin{equation} \label{eq:dissipation_inequality1}
{\cal{D}}_{iss} = -  \phi_{GB} \,  \left( \frac{\partial \psi_{GB}}{ \partial \beta} + \frac{\mu}{v_A} \right) \, \dot{\beta} + \phi_{GB} \,\boldsymbol{\sigma} :  \dot{\boldsymbol{\varepsilon}}^{diff}  - \nabla \left[ \, \mu \, \right] \cdot \vec{h}_V   \geq 0 \, ,
\end{equation}

\noindent 
which consists of chemical, mechanical, and diffusional contributions. Equation \eqref{eq:dissipation_inequality1} can be further rearranged by considering that

\begin{equation*}
\phi_{GB} \, \boldsymbol{\sigma} :  \dot{\boldsymbol{\varepsilon}}^{diff} =  \phi_{GB} \, \left( \boldsymbol{\sigma} :  \vec{n}_{GB} \otimes \vec{n}_{GB} \right)  \, \dot{\beta}  \, ,
\end{equation*}

\noindent
from which we can define $t_n =  \boldsymbol{\sigma} :  \vec{n}_{GB} \otimes \vec{n}_{GB} $ as the normal traction at GBs. It thus results that

\begin{equation} \label{eq:dissipation_inequality2}
{\cal{D}}_{iss} = \phi_{GB} \,  \left( t_n -\frac{\partial \psi_{GB}}{ \partial \beta} - \frac{\mu}{v_A} \right) \dot{\beta}  - \nabla \left[ \, \mu \, \right] \cdot \vec{h}_V   \geq 0 \, .
\end{equation}
 
\noindent
Under the assumptions of the Curie symmetry principle, fluxes and thermodynamic forces of different tensorial character do not couple. Inequality \eqref{eq:dissipation_inequality2} is then satisfied by the following conditions

\begin{equation} \label{eq:TD_restrictions2}
\phi_{GB} \, \left( t_n - \frac{\partial \psi_{GB}}{ \partial \beta} - \frac{\mu}{v_A} \right)  \dot{\beta}  \geq 0 \, , \quad   \quad  \nabla \left[ \, \mu \, \right] \cdot \vec{h}_V   \leq 0 \, .
\end{equation}

\section{Constitutive theory} \label{sec:constitutive_theory}

\subsection{Stress tensor and diffusional potential}

Both in grain boundaries and grain interiors, the Helmholtz free energy density $\psi$ is decomposed into two separate parts: a mechanical contribution $\psi^{mech}$ and a chemical contribution $\psi^{ch}$

\begin{gather*}
\psi_{bulk} \left( c_V \, , \boldsymbol{\varepsilon}^{ce}  \right) = \psi_{bulk}^{mech} \left(c_V \,,  \boldsymbol{\varepsilon}^{ce} \right) + \psi_{bulk}^{ch} \left( c_V \right) \, , \\
\nonumber \\
 \psi_{GB} \left( c_V \, , \boldsymbol{\varepsilon}^{ce} \, , \beta \right) = \psi_{GB}^{mech} \left(c_V \,,  \boldsymbol{\varepsilon}^{ce} \right) + \psi_{GB}^{ch} \left( c_V \, , \beta \right) \, .
\end{gather*}

\noindent
The mechanical part of the free energy is function of the elastic strain and it is defined as a quadratic form 
\begin{subequations}\label{eq:mechanical_free_energy}
\begin{gather} 
 \psi_{bulk}^{mech} \left( c_V \, ,  \boldsymbol{\varepsilon}^{ce} \right) = \frac{1}{2} \, \left( \boldsymbol{\varepsilon}^{ce} - \boldsymbol{\varepsilon}^{V}   \right) : \mathds{C}_{bulk} : \left( \boldsymbol{\varepsilon}^{ce} - \boldsymbol{\varepsilon}^{V}   \right) \, , \\
 \nonumber \\
\psi_{GB}^{mech} \left( c_V \, ,  \boldsymbol{\varepsilon}^{ce} \right) = \frac{1}{2} \, \left( \boldsymbol{\varepsilon}^{ce} - \boldsymbol{\varepsilon}^{V}   \right) : \mathds{C}_{GB} : \left( \boldsymbol{\varepsilon}^{ce} - \boldsymbol{\varepsilon}^{V}   \right) \, ,
\end{gather}
\end{subequations}

\noindent
where $\mathds{C}_{bulk}$ and  $\mathds{C}_{GB}$ are fourth-order elasticity tensors in bulk and GB regions.

\medskip
\noindent
The chemical part of the free energy is defined by an ideal solution model  \cite{DeHoffBook} as follows

\begin{subequations} \label{eq:chemical_free_energy}
\begin{gather}
\psi_{bulk}^{ch} \left( c_V \,  \right) = c_V \, E_V^{bulk} + R \, T \, c_L   \left[ \frac{c_V}{c_L} \, \text{ln} \left[ \frac{c_V}{c_L  }\right] + \frac{c_L  - c_V}{c_L } \,  \text{ln} \left[ \frac{c_L  - c_V}{c_L} \right] \right] \,  , \\
\nonumber \\
\psi_{GB}^{ch} \left( c_V \, , \beta \right) = c_V \, E_V^{GB} + R \, T \, c_L (\beta)   \left[ \frac{c_V}{c_L (\beta)} \, \text{ln} \left[ \frac{c_V}{c_L (\beta)  }\right] + \frac{c_L (\beta) - c_V}{c_L (\beta) } \,  \text{ln} \left[ \frac{c_L (\beta)  - c_V}{c_L (\beta)} \right] \right] \,  ,
\end{gather}
\end{subequations}

\noindent
where $E_V^{bulk}$ and $E_V^{GB}$ are energies of formation of vacancies in bulk and GB regions, while $R$ is the universal gas constant. The first part of (\ref{eq:chemical_free_energy}a) and (\ref{eq:chemical_free_energy}b) is of energetic nature, i.e. is the energy associated with one mole of vacancies in the lattice. The second part is the entropy of mixing multiplied by the absolute temperature. 
Note that, in view of equation \eqref{eq:lattice_rate}, the concentration of lattice sites $c_L$ in GB regions is not constant in general. Indeed, its evolution depends on the internal variable $\beta$ according to Eq. \eqref{eq:relationship_climb_lattice}. 

\medskip

The stress tensor $\boldsymbol{\sigma}$ and the diffusional potential $\mu$ descend from thermodynamic restrictions \eqref{eq:TD_restrictions1} that, in view of definitions \eqref{eq:mechanical_free_energy} and  \eqref{eq:chemical_free_energy}, yield

\begin{subequations}
\begin{gather}
\boldsymbol{\sigma} = \Big[ (1-\phi_{GB} ) \, \mathds{C}_{bulk} + \phi_{GB} \, \mathds{C}_{GB}   \Big] : \left(  \boldsymbol{\varepsilon}^{ce} - \boldsymbol{\varepsilon}^{V}  \right) \, ,  \label{eq:const_stress_tensor}\\
\nonumber \\
\mu = (1-\phi_{GB} ) \, E_V^{bulk} + \phi_{GB} \,E_V^{GB}  + (1-\phi_{GB} ) \, RT \,  \text{ln} \left[ \frac{c_V}{c_L   - c_V }\right] + \phi_{GB} \, RT \,   \text{ln} \left[ \frac{c_V}{c_L ( \beta )  - c_V }\right] - \omega_V \, \text{tr} \left[ \boldsymbol{\sigma} \right] \, . \label{eq:chem_potential}
\end{gather}
\end{subequations}

\noindent 
Finally, in case that  $c_V \ll c_L$, the evolution of $c_L$ in grain boundaries in Eq. \eqref{eq:chem_potential} can be neglected. With this simplifying hypothesis, the diffusion potential can be rewritten as

\begin{equation} \label{eq:const_diff_potential}
\mu = \Big[ (1-\phi_{GB} ) \, E_V^{bulk} + \phi_{GB} \,E_V^{GB}   \Big]  + \, RT \, \text{ln} \left[ \frac{c_V}{c_L }\right] - \omega_V \, \text{tr} \left[ \boldsymbol{\sigma} \right] \, .
\end{equation}

\subsection{Flux of vacancies}

The constitutive definition of $\vect{h}_V$ must satisfy the constraint reported in Eq. \eqref{eq:TD_restrictions2}. The generalized Fick's law offers a thermodynamically consistent choice for the flux of vacancies 

\begin{equation} \label{eq:generalized_ficks}
\vec{h}_V = - \boldsymbol{M}_V \, \nabla \left[ \, \mu \, \right] \, , 
\end{equation}

\noindent
where $\boldsymbol{M}_V$ is the (positive definite) second-order mobility tensor of vacancies.  By means of equation \eqref{eq:TD_restrictions_mu} it results that

\begin{gather*}
 \vec{h}_V =  - \boldsymbol{M}_V  \left( \frac{\partial \psi_{GB}^{mech}}{\partial c_V} + \frac{\partial \psi_{GB}^{ch}}{\partial c_V}  - \frac{\partial \psi_{bulk}^{mech}}{\partial c_V}  - \frac{\partial \psi_{bulk}^{ch}}{\partial c_V} \right) \nabla \left[  \phi_{GB} \right] \, + 
\\ \nonumber \\
 -  (1- \phi_{GB})  \boldsymbol{M}_V \left[ \left( \frac{\partial^2 \psi_{bulk}^{mech}}{\partial c_V^2} + \frac{\partial^2 \psi_{bulk}^{ch}}{\partial c_V^2}  \right)   \nabla \left[  c_V \right] \, +  \left( \frac{\partial^2 \psi_{bulk}^{mech}}{\partial c_V  \partial  \boldsymbol{\varepsilon}^{ce}} + \frac{\partial^2 \psi_{bulk}^{ch}}{\partial c_V \partial  \boldsymbol{\varepsilon}^{ce}}  \right)  : \nabla \left[  \boldsymbol{\varepsilon}^{ce} \right]   \right] + \\
 \nonumber \\
  -  \phi_{GB}  \boldsymbol{M}_V  \left[ \left( \frac{\partial^2 \psi_{GB}^{mech}}{\partial c_V^2} + \frac{\partial^2 \psi_{GB}^{ch}}{\partial c_V^2}  \right)   \nabla \left[ \, c_V \right]  +  \left( \frac{\partial^2 \psi_{GB}^{mech}}{\partial c_V  \partial  \boldsymbol{\varepsilon}^{ce}} + \frac{\partial^2 \psi_{GB}^{ch}}{\partial c_V  \partial  \boldsymbol{\varepsilon}^{ce}}  \right)  : \nabla \left[  \boldsymbol{\varepsilon}^{ce} \right]  +   \left( \frac{\partial^2 \psi_{GB}^{mech}}{\partial c_V  \partial  \beta} + \frac{\partial^2 \psi_{GB}^{ch}}{\partial c_V  \partial  \beta}  \right):\nabla \left[  \beta   \right]  \right]  \, .
\end{gather*}
\noindent

\noindent
To account for a different mobility of vacancies in the boundary regions, the mobility tensor is specialized as follows 

\begin{equation*} 
\boldsymbol{M}_V = ( 1 - \phi_{GB})  \, \frac{c_V}{R \, T }  \, \left( \frac{c_L  - c_V}{c_L  } \right)  \, \boldsymbol{D}_V^{bulk}  \, + \phi_{GB}  \, \frac{c_V}{R \, T } \, \left( \frac{c_L (\beta) - c_V}{c_L (\beta) } \right)  \,  \boldsymbol{D}_V^{GB} , 
\end{equation*}

\noindent
with $\boldsymbol{D}^{bulk}_V$ and $\boldsymbol{D}^{GB}_V$ referring to the diffusivity tensors of vacancies in the lattice and in GBs, respectively. In the simple case of $c_V \ll c_L$, the vacancy flux becomes

\begin{equation} \label{eq:const_vacancies_flux}
\begin{aligned}
\vec{h}_V = & - \left( E_V^{GB} - E_V^{bulk} \right) \, \boldsymbol{M}_V  \nabla \left[  \phi_{GB} \right] - ( 1 - \phi_{GB})   \, \boldsymbol{D}^{bulk}_V   \left[ \nabla\left[ c_V \right] - \frac{  \omega_V}{R \,T}   \, c_V \, \nabla\left[ \text{tr} [ \boldsymbol{\sigma}] \right]  \right] +   \\ & - \phi_{GB} \,    \boldsymbol{D}^{GB}_V  \left[  \nabla \left[ c_V \right]  -  \frac{  \omega_V}{R \,T}  \, c_V \, \nabla \left[ \text{tr} [ \boldsymbol{\sigma}] \right] \right] \, . 
 \end{aligned}
\end{equation}

\noindent
\textbf{Remark.}
The diffusivity of vacancies can be estimated from the coefficient of atomic diffusion, usually determined from experimental evidence. The flux of atoms yields (see Herring \cite{Herring1950} for details)

\begin{equation*}
\vec{h}_A = - \boldsymbol{M}_A  \, \nabla \left[ \,-  \mu \, \right] \, ,
\end{equation*}

\noindent
where $\boldsymbol{M}_A$ is the mobility of atoms. Assuming only diffusion through the lattice and $c_V \ll c_L$, the mobilities can be expressed as

\begin{equation*}
 \boldsymbol{M}_A  = \frac{\boldsymbol{D}^{bulk}_A}{R \, T} \, c_L  \, , \qquad \text{and}  \qquad \; \boldsymbol{M}_V  = \frac{\boldsymbol{D}^{bulk}_V}{R \, T} \, c_V \, ,
\end{equation*}

\noindent
being $\boldsymbol{D}^{bulk}_A$ the lattice diffusivity of atoms. In view of Eq. \eqref{eq:generalized_ficks} along with the condition $\vec{h}_A = -\vec{h}_V$,  $\boldsymbol{M}_A$ and $\boldsymbol{M}_V$ must equate, giving

\begin{equation} \label{eq:atomic_vacancy}
 \boldsymbol{D}^{bulk}_V =  \boldsymbol{D}^{bulk}_A \frac{c_L}{c_V} \, . 
\end{equation}

\noindent
The same applies for GB diffusion.

\subsection{Rate of diffusional creep}

A possible definition of the rate of diffusional creep in accordance with thermodynamic restrictions  \eqref{eq:TD_restrictions2} is 

\begin{equation} \label{eq:linear_law_climb}
\dot{\beta} = L_{GB} \, \left( t_n - \frac{\partial \psi_{GB}^{ch} }{ \partial \beta} - \frac{\mu}{v_A} \right) \, ,  
\end{equation}

\noindent
where $L_{GB}$ is a non-negative kinetic constant. The evolution of $\beta$ is then driven by a force of chemo-mechanical nature. 
Note that for $c_V \ll c_L$, $\partial \psi^{ch}_{GB} / \partial \beta \rightarrow 0$.

To further understand the physical meaning of $L_{GB}$, it is necessary to rearrange equation \eqref{eq:linear_law_climb}. The goal is to link the continuum definition of diffusional creep with dislocation based mechanisms. Following the discussion of Section \ref{subsec:mechanics_of_creep}, the microscopic aspects of diffusional creep are related to climb of GB dislocations. In this view, $\dot{\beta}$ can be expressed by adaptation of the Orowan equation to climbing dislocations as

\begin{equation} \label{eq:specification_beta}
\dot{\beta} = \overline{v}_{dis} \, \rho_m \, b_n \, ,
\end{equation}

\noindent
where $\overline{v}_{dis}$ is the average velocity of dislocations along GBs, $\rho_m$ is the density of mobile boundary dislocations, and $b_n$ is the component of Burgers' vector normal to the boundary {\footnote{The density of GB dislocations is directly linked with the nature of the particular GB and therefore it will depend on the GB missorientation and inclination. Frank-Bilby's equation \cite{Frank1953,Bilby1955} can be used as a tool to obtain the dislocation content of each grain boundary and therefore to provide a dependency of the density of mobile dislocation, $\rho_m$ with the GB geometry.}. 
Similarly, the driving force for diffusional creep can be rewritten as

\begin{equation} \label{eq:specification_force}
\left( t_n - \frac{\partial \psi_{GB}^{ch}}{ \partial \beta} - \frac{\mu}{v_A} \right) = \frac{\overline{F}_{dis}}{b_n} \, ,
\end{equation}
 
\noindent
with $\overline{F}_{dis}$ indicating the average climb force per unit length acting on GB dislocations. Owing to equations \eqref{eq:specification_beta} and \eqref{eq:specification_force}, the  rate of diffusional creep \eqref{eq:linear_law_climb} is then equivalent to

\begin{equation*}
\overline{v}_{dis} = M_{dis} \, \overline{F}_{dis}  \, , 
\end{equation*}

\noindent
where $M_{dis} = L_{GB} / \rho_m / b_n^2 $ defines the dislocation mobility in the same spirit of Ashby \cite{Ashby1969}. Therefore, the kinetic constant $L_{GB}$ can be expressed in terms of density and mobility of GB dislocations as follows

\begin{equation}
L_{GB} = M_{dis} \rho_m  b_n^2 \, .
\end{equation}

\noindent
The values of $M_{dis}$ and $\rho_m$ will depend on the nature of each GB. In the case of  $M_{dis}$, it will generally depend  on the composition and microstructure of the alloy considered through the lattice distortion caused by different alloy species or the presence of small precipitates. The value of $\rho_m$  will be linked to the geometrical definition of the grain boundary (missorientation and inclination) and might be influenced by the stress acting on it. In this work we will only consider pure metals so grain boundary dislocation mobility is determined by the kinetics of atoms rearrangement in the boundary. If such, the mobility is said $intrinsic$ or $local$ and yields \cite{Artz1983},

\begin{equation} \label{eq:intrinsic_mobility}
M_{dis}^I = \frac{C_I \, \diffusivity_A  \, b_b}{k \, T} \, , 
\end{equation}

\noindent
where $\diffusivity_A$ is the atomic diffusion, $b_b$ the Burgers' vector of GB dislocations, $k$ the Boltzmann's constant, and $C_I$ a constant of about unity.  Respect the density of mobile GB dislocations, as stated before, it will depend on the particular geometry of each grain boundary and on the stress state acting on it. A simple election of the density of GB mobile dislocation is given by 
\begin{equation} \label{eq:disl_density}
\rho_m = \frac{C_D \,t_n^2}{G^2 \, b_b^2 } \, , 
\end{equation}

\noindent
where  $C_D$ is a constant ---that will in general depend on the total density of GB dislocations and will be therefore dependent on the GB geometrical description--- and $G$ is the shear modulus.  Such a dependence of $\rho_m$ on the applied stress can be derived assuming conditions that lead to a steady state value of mobile dislocations, as proved by Kocks et al. \cite{Kocks1975}. More complex evolution laws for $\rho_m$ can also be employed. However, for the scope of the present work, the estimation of $\rho_m$ through Eq. \eqref{eq:disl_density} is considered sufficient.

\section{Summary of the governing equations} \label{sec:governing_equations}

Based on the theory developed in the previous sections, diffusional creep in polycrystalline solids involves a coupled problem of mechanics with diffusion and generation/annihilation of vacancies at grain boundaries. The molar content of vacancies is usually negligible compared to that of atoms, thus the condition $c_V \ll c_L$ is assumed from now on. Moreover, by exploiting the condition $c_V \ll c_L$, the evolution of $c_L$  can be ignored as its impact on equations \eqref{eq:const_diff_potential} and \eqref{eq:const_vacancies_flux is negligible. 
Therefore, $c_L$ is taken constant in any point of the domain.}
Governing equations are written in terms of vacancies concentration $c_V$, displacements $\vec{u}$, and GB reaction coordinate $\beta$. Field equations are defined in a spatial region $\Omega$, typically a RVE consisting of a collection of grains, and in a time interval $[t_0, t_f]$. A summary of the governing equations is reported below.

\begin{subequations} \label{eq:governing}
\begin{enumerate}

\item Transport and generation/annihilation of vacancies at GB
\begin{equation} \label{eq:gov_transp_vac}
\dot{c_V} + \text{div} \left[ \vec{h}_V  \right] = \phi_{GB} \,  s_V  \, , 
\end{equation}
\noindent
where $\vec{h}_V$ and $s_V$ are given by  \eqref{eq:const_vacancies_flux} and \eqref{eq:relationship_climb_source}, respectively.

\item Balance of forces
\begin{equation}
\text{div} \left[ \boldsymbol{\sigma} \right] = \vec{0} \, ,
\end{equation}
\noindent
with the stress tensor $\boldsymbol{\sigma}$ given by  \eqref{eq:const_stress_tensor}. 

\item Rate of diffusional creep
\begin{equation} \label{eq:gov_climb_GBdisl}
\dot{\beta} = M_{dis} \, \rho_m \, b_n^2  \, \left( t_n - \frac{\mu}{v_A}  \right) \, , 
\end{equation}
\noindent
with $\mu$ defined by \eqref{eq:const_diff_potential}.
\end{enumerate}
\end{subequations}

\medskip

To ensure the solvability of equations \eqref{eq:governing}, boundary conditions are prescribed along Neumann $\partial^N \Omega$ and Dirichlet $\partial^D \Omega$ boundaries. For the problem at hand, Neumann boundary conditions are

\begin{subequations} \label{eq:NeumanBC}
\begin{gather}
\vec{h}_V \cdot \vec{n} = \overline{h}_V  \,,  \label{eq:fluxBC}  \\ \nonumber
\\ 
 \boldsymbol{\sigma}  \vec{n} = \vec{\overline{t}} \, , \label{eq:tracBC}
\end{gather}
\end{subequations}

\noindent
while Dirichlet Boundary conditions read

\begin{gather*}
c_V = \overline{c}_V \,,   \\
\\
 \vec{u} = \vec{\overline{u}} \, .
\end{gather*}

Initial conditions are specified for concentrations of vacancies $c_V$ and for the reaction coordinate $\beta$. Assuming equilibrium conditions at initial time, the initial concentration of vacancies is function of temperature and applied mechanical pressure

\begin{equation*}
c_V^0 = c_L \, \text{exp} \left[ - \frac{ (1-\phi_{GB} ) \, E_V^{bulk} + \phi_{GB} \,E_V^{GB}   }{R \, T } + \frac{\omega_V}{R T} \,  \text{tr} \left[ \boldsymbol{\sigma} \right] \right]. 
\end{equation*}

\subsection{Diffusion controlled and interface controlled creep} \label{sec:diff_reac_creep}

Diffusional creep is triggered by the imbalance of two different forces as denoted in Eq. \eqref{eq:gov_climb_GBdisl}: (i) a mechanical-like force, i.e. $t_n$, which results from the application of a stress, (ii) a chemical-like force, i.e. $\mu / v_A$, which is proportional to the diffusional potential. Since the latter depends on $c_V$ and this volume fraction evolves according to equation \eqref{eq:gov_transp_vac}, the rate of plastic strain at GB depends on diffusion as well. Depending on the rates of the concurrent processes of vacancies diffusion and generation/annihilation, boundaries are said \textit{perfects} or \textit{imperfects}.
A quantitative estimation can be obtained as shown next. Considering vacancy diffusion (Eq. \ref{eq:gov_transp_vac}) only in the grain boundary ($\phi_{GB} = 1$) and considering isotropic diffusivity --- i.e. $\boldsymbol{D}_{V}^{GB} = \diffusivity_{V}^{GB} \boldsymbol{I}$  --- in the vacancy flux (Eq. \ref{eq:const_vacancies_flux}) leads to
\begin{equation} \label{eq:strady_state_diff}
\dot{c}_V + \text{div} \left[    \diffusivity^{GB}_V  \nabla \left[ c_V \right]  -  \frac{   \diffusivity^{GB}_V \,  \omega_V}{R \,T} \, c_V \, \nabla \left[ \text{tr} [ \boldsymbol{\sigma}] \right]  \right] =  \frac{M_{dis} \, \rho_m \, b_n^2}{v_A}  \, \left( t_n - \frac{\mu}{v_A}  \right) \, .
\end{equation}

\noindent
Introducing the following adimensional variables

\begin{equation*}
\vec{x}^* = \frac{\vec{x}}{\bar{l}} \, ,  \qquad c_V^* = \frac{c_V}{\bar{c}} \, , \qquad \boldsymbol{\sigma}^* = \frac{\boldsymbol{\sigma}}{\bar{\sigma}} \, ,  \qquad \mu^* = \frac{\mu \, \bar{c}}{\bar{\sigma}} \, ,
\end{equation*}

\noindent
where $\bar{l}$ is a reference length, $\bar{c}$ is a reference concentration, and $\bar{\sigma}$ a reference stress, equation
\eqref{eq:strady_state_diff} is equivalent to 

\begin{equation} \label{eq:nondim_diff}
 \dot{c}^*_V + \frac{\diffusivity^{GB}_V}{\bar{l}^2 }  \, \text{div$^*$}\left[  \nabla^*\left[ c_V^* \right]  -  \frac{ \omega_V \, \bar{\sigma} }{R \,T} \, c_V^* \,  \nabla^* \left[ \text{tr} [ \boldsymbol{\sigma^*}] \right]  \right] =  \frac{M_{dis} \, \rho_m \, b_n^2 \, \bar{\sigma}}{   v_A \,  \bar{c}}  \, \left( t_n^* - \frac{\mu^*}{v_A^*}  \right) \, ,
\end{equation}

\noindent
where $v_A^*=\bar{c} \, v_A$ and $\omega_V \bar{\sigma} / (R \,T)$ are adimensional constants. From equation \eqref{eq:nondim_diff}, $\tau_D = \bar{l}^2  / \diffusivity^{GB}_V$ identifies the characteristic time controlling the diffusional process, while $ \tau_S = v_A \, \bar{c} / M_{dis} \, \rho_m \, b_n^2 \, \bar{\sigma} $ is the characteristic time of the GB reaction. Therefore, the following adimensional magnitude

\begin{equation} \label{eq:creep_regime}
\epsilon= \frac{\tau_D}{\tau_S} = \frac{\bar{l}^2 \, M_{dis} \, \rho_m \, b_n^2 \, \bar{\sigma}}{\diffusivity^{GB}_V \, v_A \, \bar{c}  } \, ,
\end{equation} 

\noindent
controls the ratio between the rate of the two processes. On the one hand, if $\epsilon \gg 1$ diffusion is much slower than vacancies generation/annihilation, highlighting the presence of perfect boundaries: the creep rate is \textit{diffusion controlled}. On the other hand, for $\epsilon \ll 1$ operation of vacancies sink/source is the rate limiting process and creep is said \textit{reaction} or \textit{interface controlled}. 

\medskip 

\subsubsection{Diffusion controlled creep}

When $\epsilon \gg 1$, the time required for \eqref{eq:gov_climb_GBdisl} to reach equilibrium is much smaller than the time scale of diffusion. In such event, it can be assumed that the kinetics of GB dislocation is so fast that its driving force equilibrates instantaneously. The diffusional potential in the boundaries is then given as a function of the applied stress as $\mu^{eq} = t_n \, v_A$. It follows that the equilibrium concentration of vacancies at GB yields 

\begin{equation} \label{eq:vacancy_eq_perfect}
c_V^{eq} = \hat{c}_V \, \text{exp} \left[ \frac{t_n \, v_A}{R \, T}\right] \, ,
\end{equation}

\noindent
being $\hat{c}_V =  c_L \, \text{exp} (-E_f^V / R \, T + \omega_V / RT \text{tr}[\boldsymbol{\sigma}])$ the reference-state concentration of vacancies. Boundaries with positive applied $t_n$ are sources of vacancies, while boundaries with negative $t_n$ act as sinks. The creep rate is then controlled only by the kinetics of diffusion of vacancies between sinks and sources. Note that the boundary values of $\mu$ and $c_V$ agree with theories of diffusion controlled creep, see Herring \cite{Herring1950} and Coble \cite{Coble1963} for instance. 

\subsubsection{Interface controlled creep}
When $\epsilon \ll 1$, diffusion is so fast that accumulation of vacancies at grain boundaries is prevented. The distribution of the diffusional potential is given by solving $\nabla \left[ \, \mu \, \right] = \vec{0}$, without resorting to eq. \eqref{eq:gov_transp_vac}. Creep rate is then controlled by Eq. \eqref{eq:gov_climb_GBdisl} solely.

\section{Numerical examples} \label{sec:numerical_ex}

\begin{figure}[htbp]
\centering
		\begin{tabular}{rl}
		\subfigure[]{ \includegraphics[width=7.5cm]{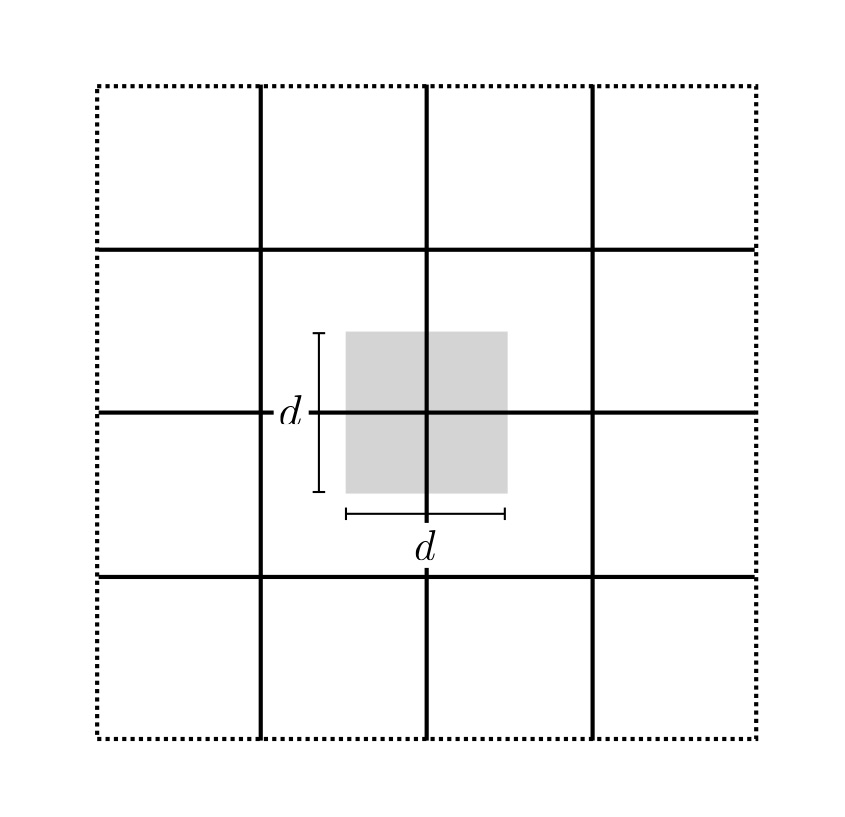}} 
		&
		\subfigure[]{ \includegraphics[width=7cm]{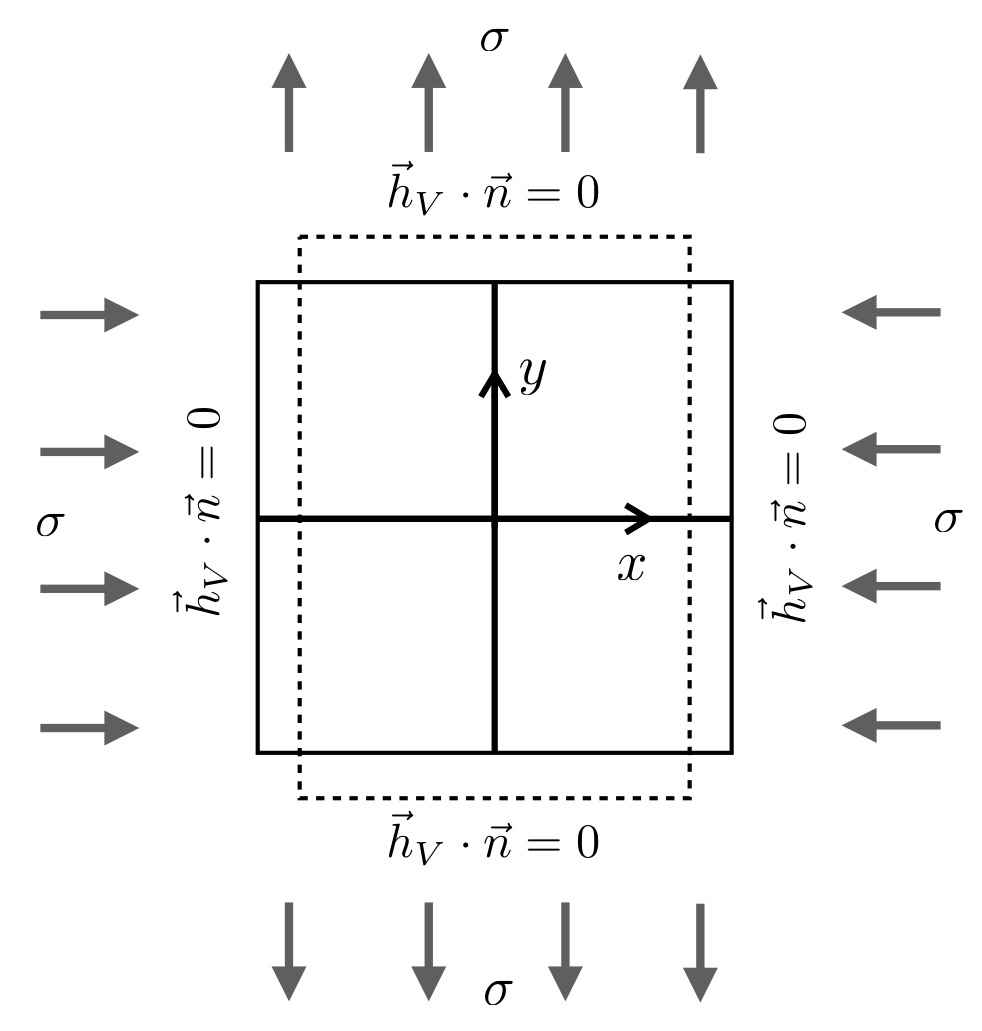}} 
		\end{tabular}
\caption{\em{Idealized solid microstructure made up of a regular array of square grains (a). Geometry and boundary conditions used in the simulations (b).}}
\label{fig:geometry}
\end{figure}

\subsection{Geometry and assumptions} 

The aim of this section is to analyze diffusional creep of polycrystalline solids by means of numerical analyses. To this end, a 2D plane-strain regular array of square grains of size $d$ is used as solid microstructure, as shown in Figure \ref{fig:geometry}a. Grain boundaries will correspond to pure tilt GB
The solid is deformed by applying a constant macroscopic shear stress $\sigma  \left( - \vec{e}_x \otimes  \vec{e}_x + \vec{e}_y  \otimes  \vec{e}_y \right)$. For simplicity, stiffness in the bulk and in GBs are assumed equal $(\mathds{C}_{bulk} = \mathds{C}_{GB})$ and isotropic. Eq. \eqref{eq:const_stress_tensor} reduces to 

\begin{equation}
\boldsymbol{\sigma} = \frac{2 \, G (1+ \nu) }{3 \, (1- 2 \nu)} \, \text{tr} \left[  \boldsymbol{\varepsilon}^{ce} -  \boldsymbol{\varepsilon}^{V} \right] \boldsymbol{I} + 2 \, G  \,\text{dev} \left[ \boldsymbol{\varepsilon}^{ce} \right] \, ,
\end{equation}

\noindent 
where $G$ is the shear modulus and $\nu$ the Poisson's coefficient. The computational domain can, therefore, be limited to the one depicted in Fig. \ref{fig:geometry}b. To account for the mechanical influence of adjacent grains, external boundaries are constrained to remain flat throughout the analysis. In addition, the normal component of the flux of vacancies at external boundaries is set zero because of symmetry. The solid is assumed to be stress-free at initial time $t_0 = 0 $ s. Assuming $E_V^{bulk} = E_V^{GB} = E_V$, the initial concentration of vacancies yields

\begin{equation*}
c_V^0 = c_L \, \text{exp} \left[ - \frac{E_V}{R \, T } \right]. 
\end{equation*}

Due to the model symmetry, horizontal and vertical GB would have different total number of GB dislocations. However, we have considered for simplicity that the mobile dislocation density,  $\rho_m$, on all the GB follows the same equation neglecting the possible effect of GB missorientation. The density of mobile dislocations is then simply computed as $\rho_m = \sigma^2 / G ^2  /  b_n^2$.  Following Artz et al. \cite{Artz1983}, $b_b = b / 3$ and $b_b / b_n = \sqrt{2}$, where $b$ is the Burgers' vector of lattice dislocations. The concentration of lattice sites is taken as the inverse of the molar volume, i.e. $c_L = 1/v_A$. The effect of the eigenstrain $\boldsymbol{\varepsilon}^V$ is neglected by taking $\omega_V = 0$ $\text{m}^3$/mol.

The diffusivity tensors are specialized as follows 

\begin{equation} \label{eq:mobility_tensor}
 \boldsymbol{D}_V^{bulk}  = \diffusivity^{bulk}_V  \, \boldsymbol{I}   \qquad \text{and} \qquad  \boldsymbol{D}_V^{GB} = \diffusivity^{GB}_V\left( \boldsymbol{I} - \vec{n}_{GB} \otimes \vec{n}_{GB} \right) \, , 
\end{equation}

\noindent
with $\diffusivity^{bulk}_V$ and $\diffusivity^{GB}_V$ referring to the diffusion coefficient of vacancies in the lattice and in grain boundaries, respectively. Diffusion is considered isotropic in the grain interior, while in the boundaries is not. This is assumed in order to prevent diffusion of vacancies across different grains. Diffusivity of vacancies are derived from atomic diffusivity according to Eq. \eqref{eq:atomic_vacancy}. Such parameters are usually expressed as Arrhenius laws \cite{FrostAshby1982}

\begin{equation*}
\diffusivity^{bulk}_A  = D_0^{bulk} \, \text{exp} \left[ \frac{- Q_a^{bulk}}{R \, T} \right] \, , \qquad
\diffusivity^{GB}_A = D_0^{GB} \, \text{exp} \left[ \frac{- Q_a^{GB}}{R \, T} \right] \, ,
\end{equation*}

\noindent
where $D_0^{bulk}$ and $D_0^{GB}$ are pre-exponential factors, while $Q_a^{bulk}$ and $Q_a^{GB}$ are activation energies for atomic diffusion. Similarly, the shear modulus yields

\begin{equation*}
G = G_0 + G_0 \, \frac{T-300}{T_M} \,  \frac{T_M}{G_0} \frac{\text{d} G }{\text{d} T } \, ,
\end{equation*}

\noindent 
where $G_0$ is the shear modulus at 300 K, $T_M$ the melting temperature, and $ \frac{T_M}{G_0} \frac{\text{d} G }{\text{d} T }$ the coefficient of temperature dependence of $G$. Material parameters representative of pure copper are used in the simulations that follow. Their numerical values are listed in Table \ref{tab:matpar}.  

Governing equations are solved numerically through the Finite Element Method by implementing the system of equations \eqref{eq:governing} in the open-source computing platform \textit{FEniCS} \cite{FenicsProject}. Further details of the numerical implementation are reported in Appendix \ref{app:num_implementation}.

\begin{table}
\centering
\begin{tabular}{  l c c c c}  

\multicolumn{1}{c}{\textbf{ Material Parameters }}  &
\multicolumn{1}{c}{\textbf{ }}  &
\multicolumn{1}{c}{\textbf{ }}  &
\multicolumn{1}{c}{\textbf{}}   & 
\multicolumn{1}{c}{\text{ Ref.}}   \\ 
\hline
\\
Pre-exponential bulk diffusion & \hspace{0.3cm}  $D_0^{bulk}  \hspace{0.3cm} $   & \hspace{0.3cm} $2.0 \times 10^{-5}$   \hspace{0.3cm}  & \hspace{0.3cm} $\rm{m}^2 / \rm{s}$  \hspace{0.3cm} &   \cite{FrostAshby1982}\\
\\
Pre-exponential GB diffusion & \hspace{0.3cm}  $D_0^{GB}  \hspace{0.3cm} $   & \hspace{0.3cm} $1.0 \times 10^{-7}$   \hspace{0.3cm}  & \hspace{0.3cm} $\rm{m}^2 / \rm{s}$  \hspace{0.3cm} &   \small{This study}\\
\\
Activation energy for bulk diffusion & \hspace{0.3cm}  $Q_a^{bulk}  \hspace{0.3cm} $   & \hspace{0.3cm} $1.97 \times 10^{5}$   \hspace{0.3cm}  & \hspace{0.3cm} $\rm{J} / \rm{mol}$  \hspace{0.3cm} &   \cite{FrostAshby1982}\\
\\
Activation energy for GB diffusion & \hspace{0.3cm}  $Q_a^{GB}  \hspace{0.3cm} $   & \hspace{0.3cm} $1.04 \times 10^{5}$   \hspace{0.3cm}  & \hspace{0.3cm} $\rm{J} / \rm{mol}$  \hspace{0.3cm} &   \cite{FrostAshby1982}\\
\\
Molar volume of atoms  & \hspace{0.3cm}  $v_{A}  \hspace{0.3cm} $   & \hspace{0.3cm} $7.1 \times 10^{-6}$   \hspace{0.3cm}  & \hspace{0.3cm} $\rm{m}^3 / \rm{mol}$  \hspace{0.3cm} &   \cite{FrostAshby1982}\\
\\
Vacancy formation energy  & \hspace{0.3cm}  $E_{V}  \hspace{0.3cm} $   & \hspace{0.3cm} $1.225 \times 10^{5}$   \hspace{0.3cm}  & \hspace{0.3cm} $\rm{J} / \rm{mol}$  \hspace{0.3cm} &   \cite{Villani2015}\\
\\
Melting temperature & \hspace{0.3cm}  $T_M  \hspace{0.3cm} $   & \hspace{0.3cm} $1356 $   \hspace{0.3cm}  & \hspace{0.3cm} $\rm{K}$  \hspace{0.3cm} &   \cite{FrostAshby1982}\\
\\
Burgers' vector of lattice dislocations  & \hspace{0.3cm}  $b  \hspace{0.3cm} $   & \hspace{0.3cm} $2.56 $   \hspace{0.3cm}  & \hspace{0.3cm}  \AA  \hspace{0.3cm} &   \cite{FrostAshby1982}\\
\\
Shear modulus at 300 K & \hspace{0.3cm}  $G_0  \hspace{0.3cm} $   & \hspace{0.3cm} $42.1$   \hspace{0.3cm}  & \hspace{0.3cm} $\rm{GPa}$  \hspace{0.3cm} &   \cite{FrostAshby1982}\\
\\
Temperature dependence of G & \hspace{0.3cm}  $  \frac{T_M}{G_0} \frac{\text{d} G }{\text{d} T }  \hspace{0.3cm} $   & \hspace{0.3cm} $-0.54$   \hspace{0.3cm}  & \hspace{0.3cm} $- $  \hspace{0.3cm} &   \cite{FrostAshby1982}\\
\\
Poisson's coefficient & \hspace{0.3cm}  $  \nu  \hspace{0.3cm} $   & \hspace{0.3cm} $0.285$   \hspace{0.3cm}  & \hspace{0.3cm} $ -  $  \hspace{0.3cm} &   \cite{FrostAshby1982}\\
\\
\hline
\end{tabular} 
\captionof{table}{Material parameters representative of pure copper adopted for the numerical simulations.}
\label{tab:matpar}
\end{table}

\subsection{Steady-state and transient evolution of diffusional creep} \label{subsec:case1}

In this section we simulate the response of a polycrystal of grain size $d=100 \, \mu$m subjected to shear stress $\sigma=10 \, \text{MPa}$ at temperature $T=900$ K (homologous temperature of 66\%). The load is applied instantaneously after the initial time $t=0$ s and is kept constant throughout the analysis. Grain boundaries are defined through the GB indicator function introduced in Eq. \eqref{eq:GB_indicator_fun}. Following Villani et al. \cite{Villani2015}, a constant GB thickness $d_{GB} = 4 \, \mu$m is selected as a compromise between reality and tractable computations. 
Note that the selected grain boundary thickness is several orders of magnitude greater than reality ($\sim 0.5$ nm). Therefore, to compensate this model limitation, the GB diffusivity was estimated \textit{ad hoc}, as to match numerical results with analytical formula by Coble \cite{Coble1963}. Accordingly, a pre-exponential factor $D_0^{GB} = 1 \times 10^{-7} \, \text{m}^2 / \text{s}$ has been assigned.

\begin{figure}[htbp]
\centering
		\begin{tabular}{cc}
		\subfigure[]{ \includegraphics[height=5.75cm]{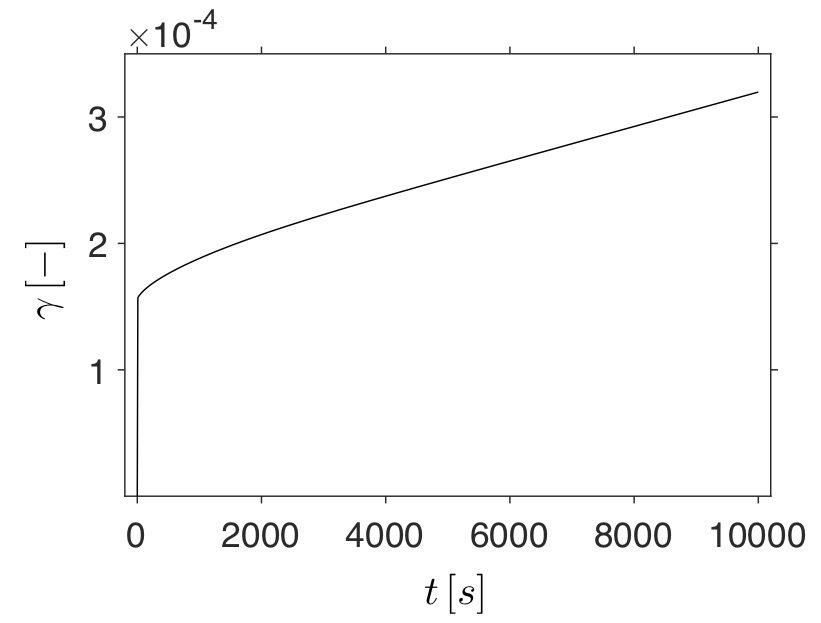} }
		&
		\subfigure[]{ \includegraphics[height=5.75cm]{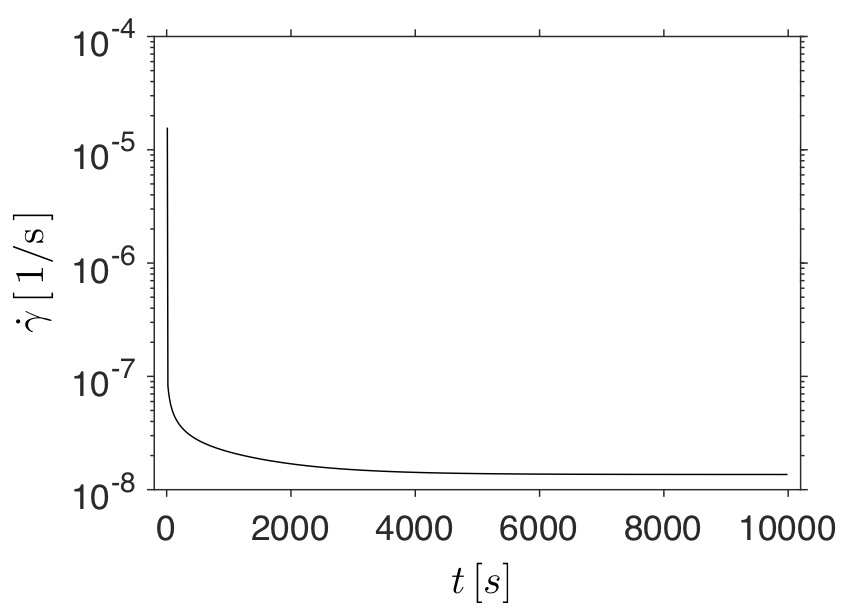}}
		\end{tabular}
\caption{\em{Plot of the average shear strain (a) and strain rate  (b) against time for a polycrystal of grain size $d=100 \, \mu$m. The solid is subjected to shear stress $\sigma =  10$ MPa at temperature $T=900$ K. The average shear strain is computed as  $\gamma = 1 / A_\Omega\,   \int_\Omega (\varepsilon_{22} -  \varepsilon_{11})/2 \, \rm{d}A$, with $A_\Omega$ denoting the RVE area.}}
\label{fig:strainVStime}
\end{figure}

Figure \ref{fig:strainVStime} shows the evolution of the average shear strain and strain rate against time. The strain grows non-linearly in time, right after the application of the load, with decreasing rate as time advances. A steady-state condition is reached after about $5000$ s, for which it results a constant strain rate $\dot{\gamma} = 1.35 \times 10^{-8}$ 1/s.

The evolution  of vacancy concentration and stress component $\sigma_{22}$ at grain boundary $y=0$ is depicted in Fig \ref{fig:boundary_intrinsic}. It results that $c_V$ is in equilibrium with the applied normal stress $\sigma_{22}$ according to Eq. \eqref{eq:vacancy_eq_perfect}. Therefore, for the considered grain size $d$, applied stress $\sigma$, and dislocation mobility, boundaries behave like perfect sinks/sources for vacancies. The stress relaxes at GB junction ($x=0$), while attains a maximum at $x=\pm 0.5$. This happens as diffusion, and then creep rate, is faster at GB junctions where concentration gradients are higher. Similarly,  $c_V > c_V^0$  everywhere in the horizontal boundary except at the quadruple junction where stress is zero. 

\begin{figure}[htbp]
\centering
		\begin{tabular}{cc}
		\subfigure[]{ \includegraphics[height=5.5cm]{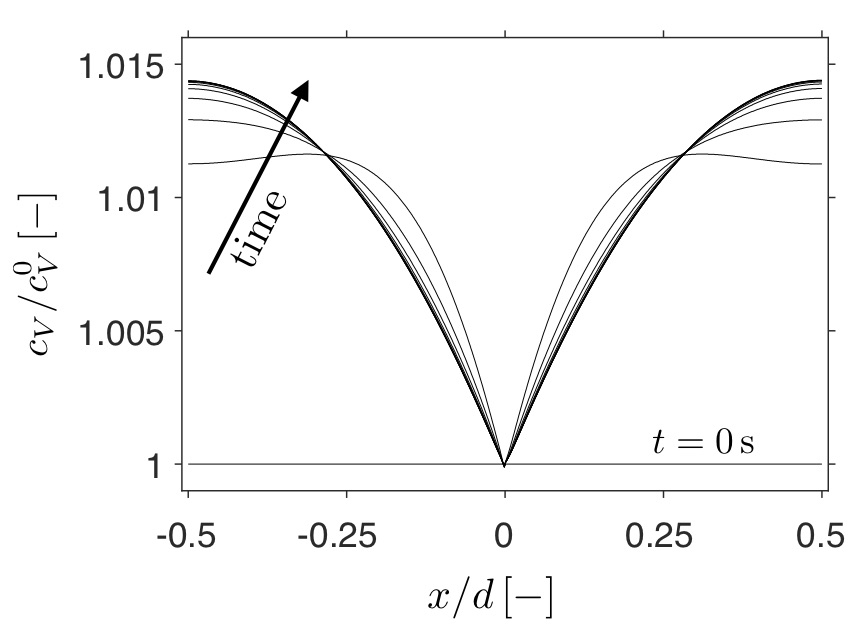}} 
		&
		\subfigure[]{ \includegraphics[height=5.5cm]{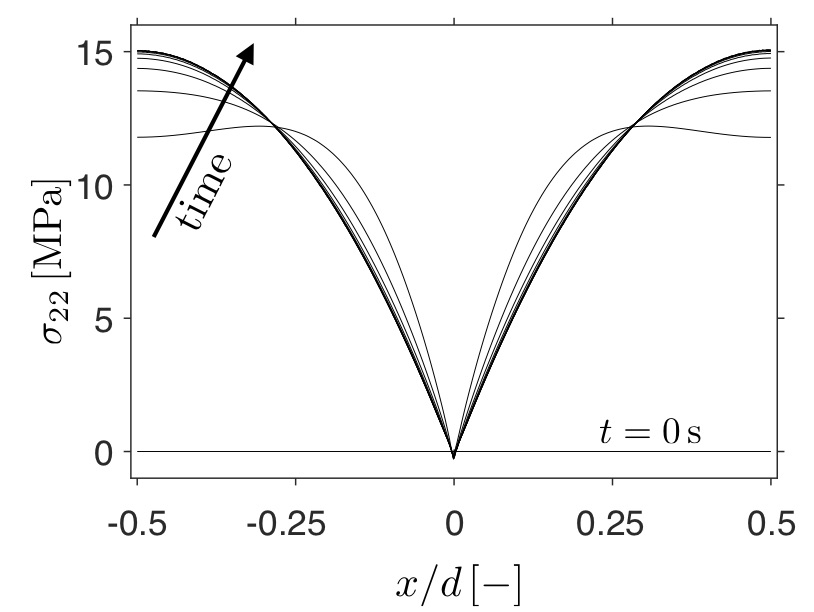}} 
		\end{tabular}
\caption{\em{Normalized vacancy concentration (a) and normal stress (b) in the horizontal GB as a function of $x$ coordinate at intervals of 1000 s. Analyses are performed on a polycrystal of grain size $d=100 \, \mu$m subjected to shear stress $\sigma = 10$ MPa and $T=900$ K.}}
\label{fig:boundary_intrinsic}
\end{figure}

The distribution of vacancies, stress and plastic strain rate are represented in Fig. \ref{fig:contours} at steady-state. Vacancies accumulate in the horizontal GB, where normal stress is positive, and deplete in the vertical one, where stress in negative. The average rate of generation of vacancies equates that of absorption, so that the overall content of point defects does not change as deformation proceeds. Similarly, the stress concentrates in grain interiors while relaxes at GB junctions, where the rate of plastic strain is maximum. 

\begin{figure}[htbp]
\centering
		\begin{tabular}{lcr}
		\subfigure[]{ \includegraphics[height=6.cm]{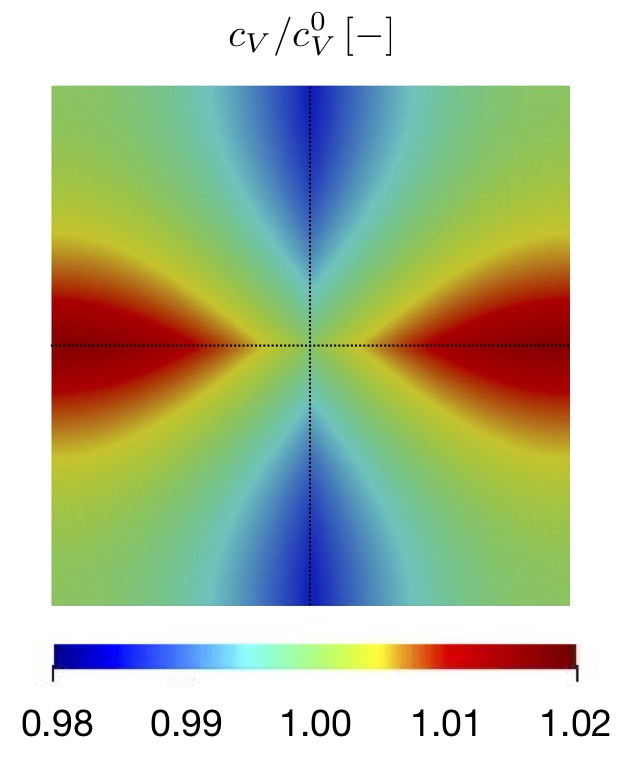}} 
		&
		\subfigure[]{ \includegraphics[height=6.cm]{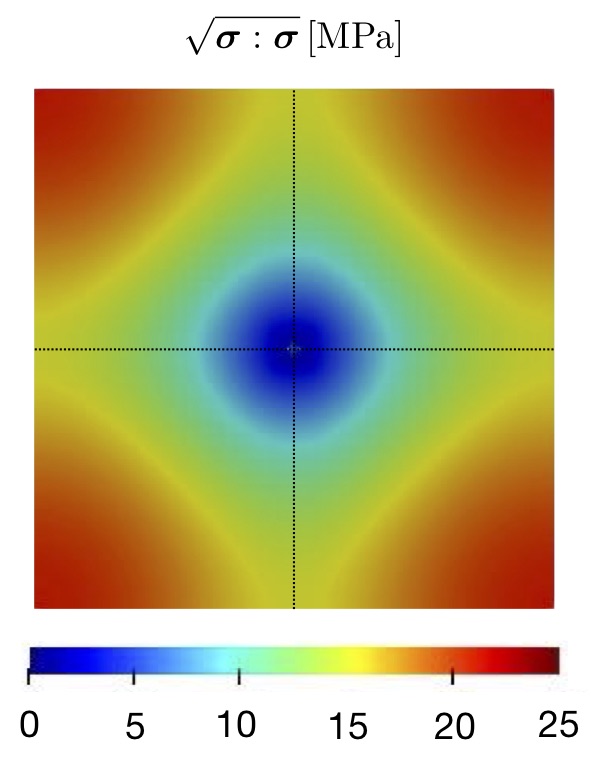}} 
		&
		\subfigure[]{ \includegraphics[height=6.cm]{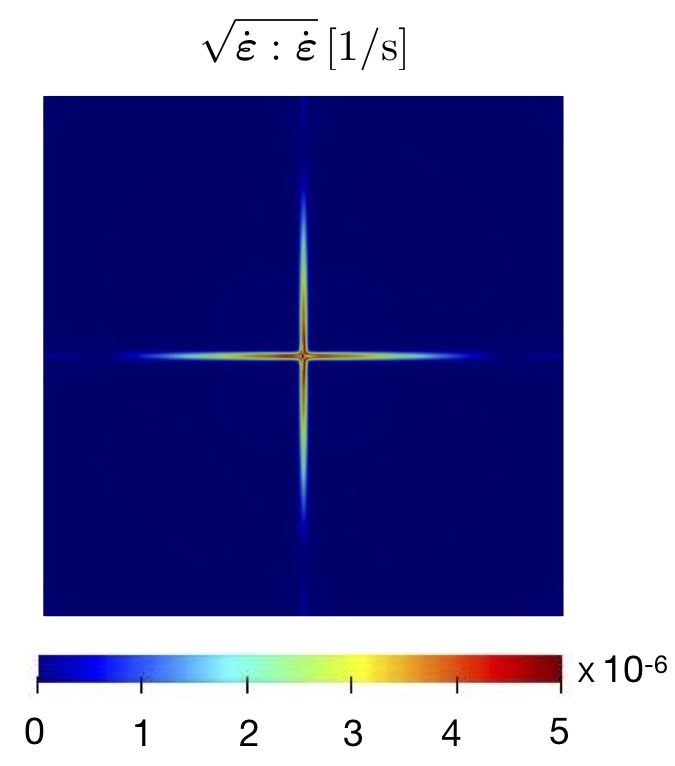}} 
		\end{tabular}
\caption{\em{Contour plots of (a) normalized vacancy concentration, (b) stress norm, and (c) norm of strain rate at time  $t=10000$ s. Results are obtained for a polycrystal of grain size $d=100 \, \mu$m subjected to shear stress $\sigma = 10$ MPa at temperature $T=900$ K.}}
\label{fig:contours}
\end{figure}

\subsection{The effect of GB dislocation mobility on diffusional creep}

We now consider the effect of the GB dislocation mobility by simulating the same case study of Section \ref{subsec:case1} with different mobilities. To show the transition from perfect to imperfect grain-boundaries, the dislocation mobility is defined as $M_{dis}=M_{dis}^{I}/ \alpha$, with $\alpha$ ranging from 1 to 0.001. The resulting creep curves are depicted in Figure \ref{fig:strainVSmobilities}. As predictable, the lower the dislocation mobility, the lower the strain rate attained at steady state. In addition, $M_{dis}$ impacts on the transient response: for low dislocation mobility a steady-state condition is achieved almost immediately after loading. 

\begin{figure}[htbp]
\centering
		\begin{tabular}{cc}
		\subfigure[]{ \includegraphics[height=5.75cm]{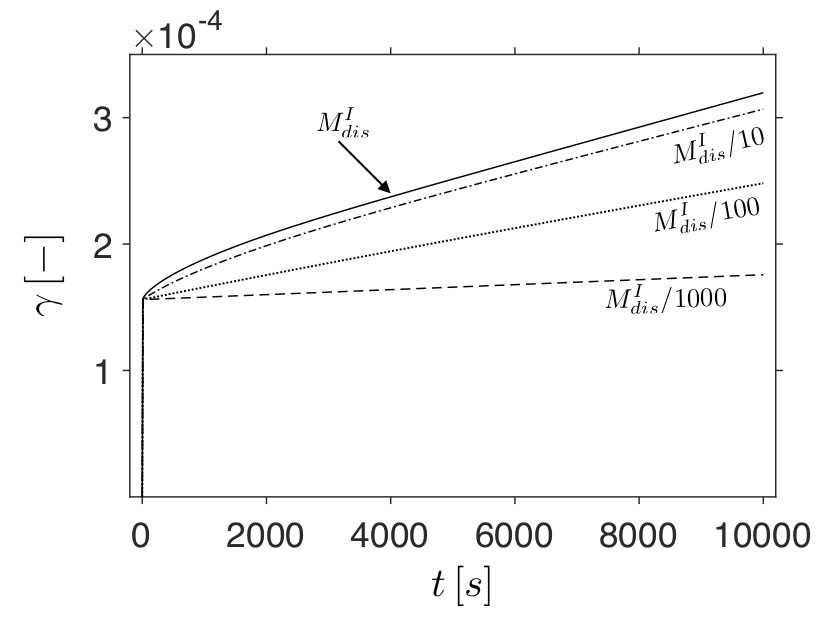}} 
		&
		\subfigure[]{ \includegraphics[height=5.75cm]{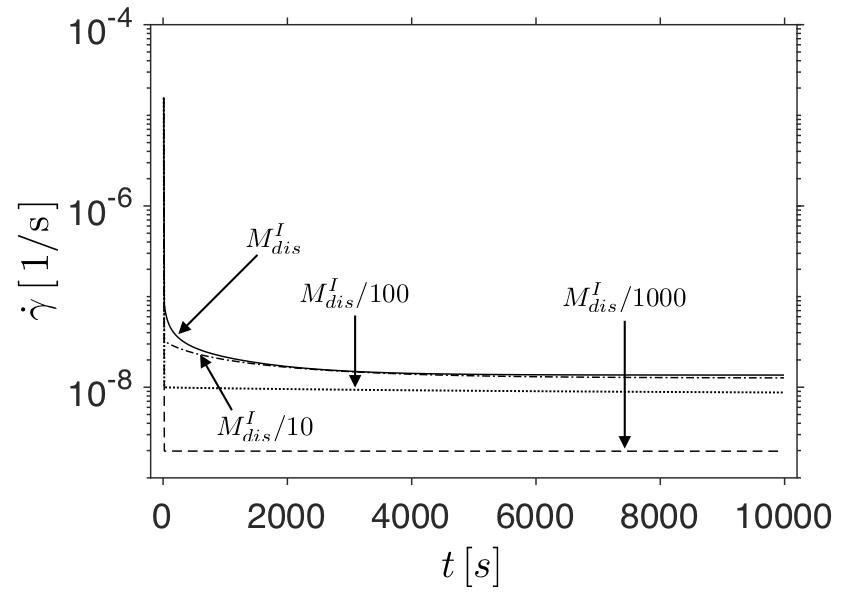}} 
		\end{tabular}
\caption{\em{Comparison between the simulated evolution of average shear strain (a) and strain rate (b) obtained for different assigned GB dislocation mobilities. Analyses are performed on a polycrystal with $d=100 \, \mu$m  subjected to a shear stress $\sigma =  10$ MPa at temperature $T=900$ K. The average shear strain is computed as  $\gamma = 1 / A_\Omega\,   \int_\Omega (\varepsilon_{22} -  \varepsilon_{11})/2 \, \rm{d}A$, with $A_\Omega$ denoting the RVE area. }}
\label{fig:strainVSmobilities}
\end{figure}


Figure \ref{fig:evo_mobilities} gathers stress and vacancy distribution in the horizontal grain boundary at steady state for different mobilities. In both graphs, the extreme cases of diffusion controlled and interface controlled creep are clearly shown. On the one hand, for $M_{dis}=M_{dis}^I$, creep is diffusion controlled, $c_V$ and $\sigma_{22}$ attains a non uniform distribution in the GB as discussed previously. On the other hand, for $M_{dis}=M_{dis}^I / 1000$, creep is reaction controlled: the kinetics of diffusion prevent vacancies from accumulating or depleting in GB regions. In such a case, the creep rate $\dot{\beta}$ is uniform on grain boundaries inducing a uniform stress which does not relax at GB junctions.  

\begin{figure}[htbp]
\centering
		\begin{tabular}{cc}
		\subfigure[]{ \includegraphics[height=5.5cm]{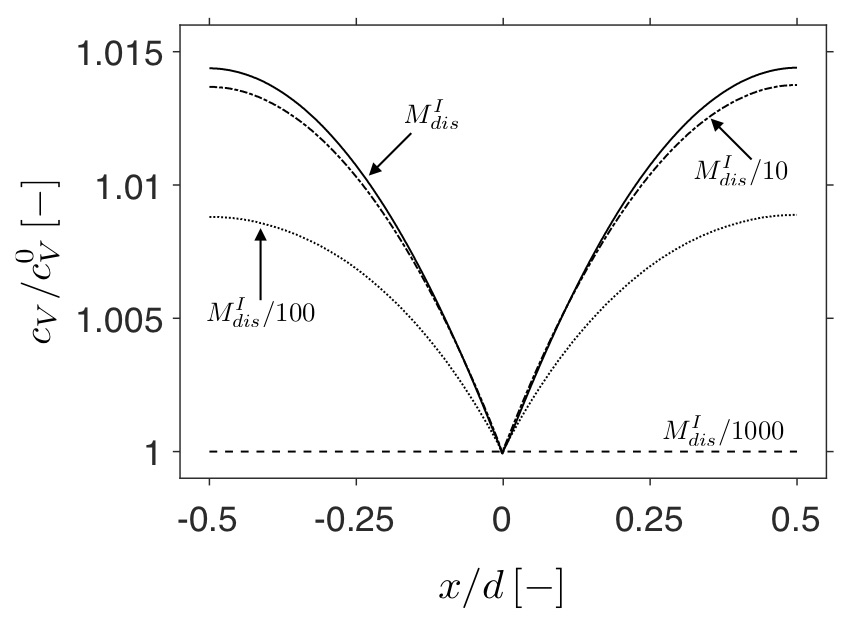}} 
		&
		\subfigure[]{ \includegraphics[height=5.5cm]{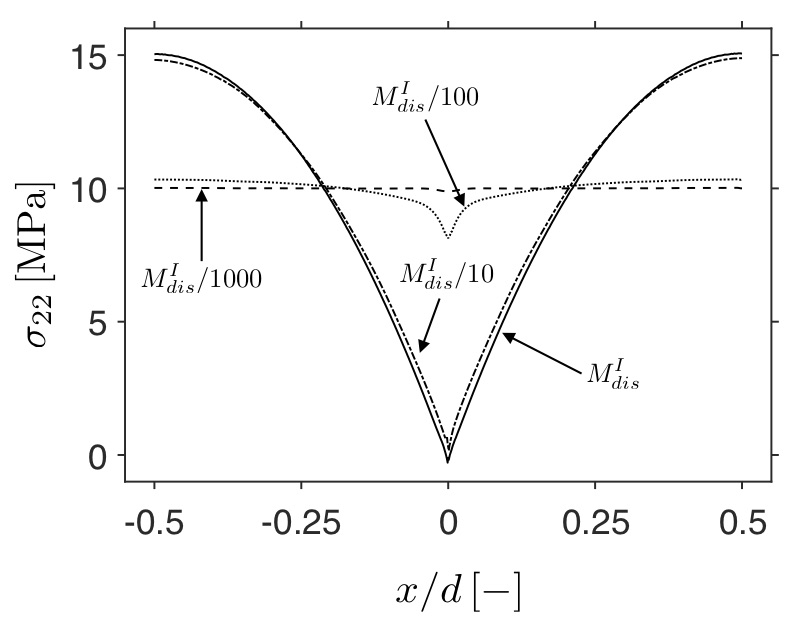}} 
		\end{tabular}
\caption{\em{Vacancy concentration (a) and normal stress (b) in the horizontal GB as a function of $x$ coordinate at steady state (t=10000 s) for varying GB dislocation mobility. Results are obtained for a polycrystal of grain size $d=100 \, \mu$m subjected to shear stress $\sigma = 10$ MPa at $T=900$ K. }}
\label{fig:evo_mobilities}
\end{figure}

For intermediate mobilities, i.e. $M_{dis}^{I}/10$ and  $M_{dis}^{I}/100$, results are in-between the extreme cases of perfect and imperfect boundaries. Note that the chemo-mechanical response for $M_{dis}^{I}/10$ does not differ significantly from the one of intrinsic mobility. Indeed, the GB reaction kinetics depends little on dislocation mobility if boundaries are nearly perfect.

\subsection{Stress and grain size dependence}

In this section, we investigate the impact of grain size and applied stress on the steady-state creep rate. Grain size dependence is studied by simulating creep tests of polycrystals with $d$ ranging from 20 to 200 microns  and applied stress $\sigma=10$ MPa. The role of stress is analyzed on a polycrystal of grain size $100 \, \mu$m for $\sigma = 1 \div 10$ MPa. Stress and grain size dependence is evaluated for high and low GB dislocation mobility, i.e. $M_{dis}^{I}$ and $M_{dis}^{I}/1000$ respectively. 

Figure \ref{fig:grain_stress_boundary} presents the numerical predictions at $T=900$ K for the  material parameters listed in Table \ref{tab:matpar}. Results are compared with the analytical formula by Coble, for which the steady-state creep rate yields

\begin{equation} \label{eq:Coble}
\dot{\gamma}_C = \frac{150 \, \sigma \, \delta \diffusivity_A^{GB}  \, v_A}{d^3 \, R \, T }, 
\end{equation}

\noindent
where  $\delta \diffusivity_A^{GB}$ is the boundary thickness times the diffusion coefficient for mass transport in the grain boundary.
For pure copper, $\delta \diffusivity_A^{GB}=2.0 \times 10^{-15} \, \text{exp} ( - 1.97 \times 10^{ 5 } / R / T ) \, \rm{m^3} / \rm{s}$  \cite{FrostAshby1982}. Numerical results with $M_{dis}^{I}$ are in agreement with Coble's formula, suggesting that in the range of selected grain size and loading stress, creep is diffusion controlled. Slope in Fig. \ref{fig:grain_stress_boundary}a is $\sim -3$ as mass transport is dominated by GB diffusion. Indeed, at the considered temperature $\diffusivity_A^{GB} \gg \diffusivity_A^{bulk}$. 

The stress and grain size dependence changes drastically with decreasing GB dislocation mobility. Assuming $M_{dis}^{I}/1000$, $\dot{\gamma} \propto \sigma^{3}/d$. This confirms that for low mobility, the GB reaction kinetics is no longer limited by diffusion. Creep is then interface controlled. The strain rate is proportional to the applied stress and GB dislocation density, as shown in Eq. \eqref{eq:gov_climb_GBdisl}. Accordingly, a slope 3 is obtained in Fig. \ref{fig:grain_stress_boundary}b since $\rho_m$ is taken proportional to the  square of the applied stress.

\begin{figure}[htbp]
\centering
		\begin{tabular}{cc}
		\subfigure[]{ \includegraphics[height=5.5cm]{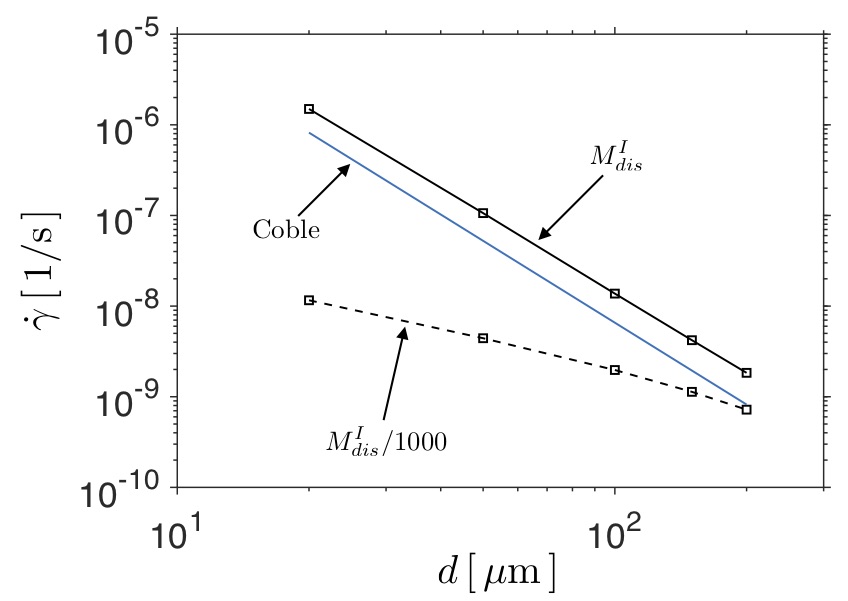}} 
		&
		\subfigure[]{ \includegraphics[height=5.5cm]{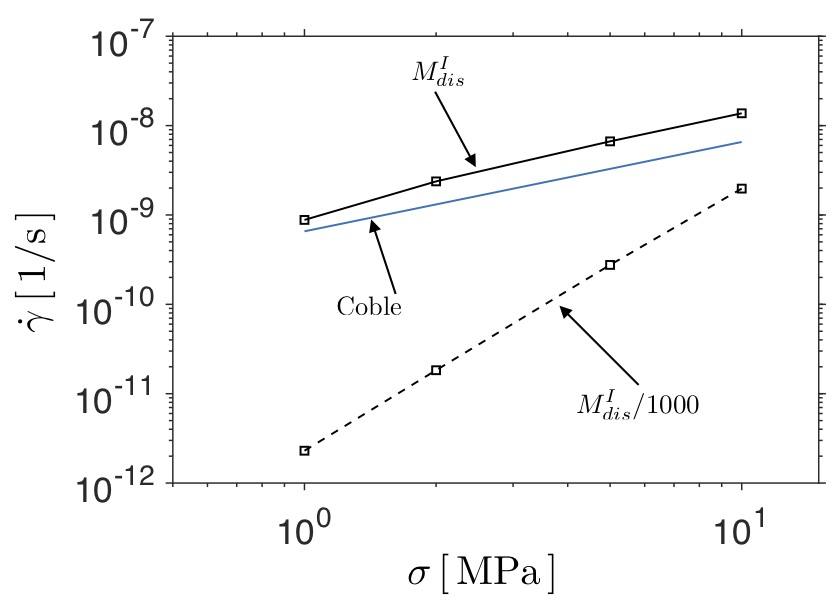}} 
		\end{tabular}
\caption{\em{ Log-log plot of steady-state creep rate against grain size (a) and against applied stress (b) at temperature $T=900$ K. Numerical predictions with different mobilities are compared with the analytical formula by Coble Eq. (\ref{eq:Coble}). Squared marker identifies individual results obtained for the simulated grain size and applied stress. }}
\label{fig:grain_stress_boundary}
\end{figure}

Grain size and stress dependence on creep rate is also studied when diffusion takes place through grain interiors. In such a case, the GB diffusivity is taken equal to the one prescribed in the lattice, i.e.  $\diffusivity_A^{GB} = \diffusivity_A^{bulk}$. Simulations have been performed in the same way as described above, but with temperature $T=1100$ K. Figure \ref{fig:grain_stress_bulk} depicts the present numerical outcomes along with the analytical steady-state creep rate estimated by Herring \cite{Herring1950}

\begin{equation} \label{eq:Herring}
\dot{\gamma}_H = \frac{8 \, \sigma \,  \diffusivity_A^{GB}  \, v_A}{d^2 \, R \, T } \, .
\end{equation}

\noindent
Similarly to the case of dominating GB diffusion, Fig. \ref{fig:grain_stress_bulk}a shows that a transition from high to low dislocation mobility induces a change in the dependence of creep rate on grain size. On the one hand, for $M_{dis}^I$, the numerical predictions are similar to Herring's formula. However, the simulated creep slope is slightly less than the one predicted by Herring, suggesting that creep is not completely diffusion controlled. On the other hand, for low mobility slope is $1$ and creep is reaction controlled as described previously.

\begin{figure}[htbp]
\centering
		\begin{tabular}{cc}
		\subfigure[]{ \includegraphics[height=5.5cm]{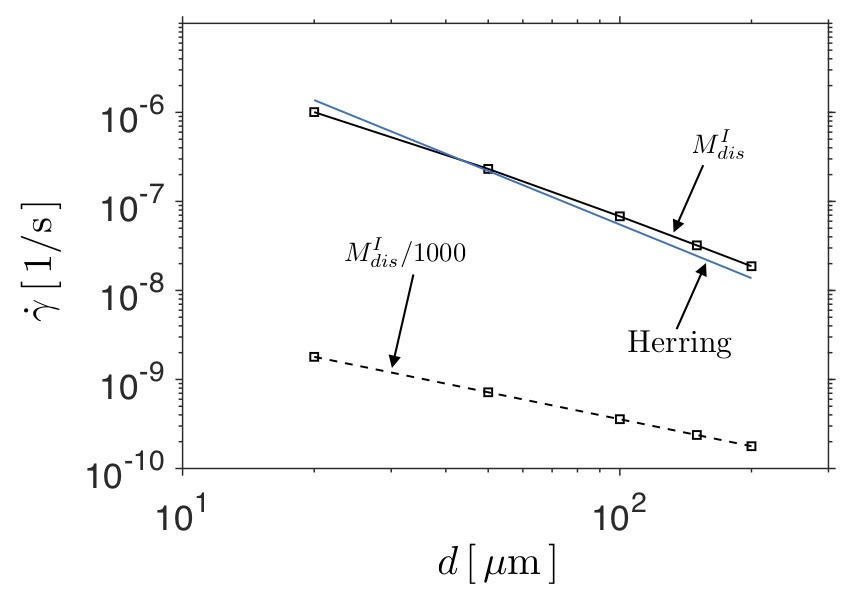}} 
		&
		\subfigure[]{ \includegraphics[height=5.5cm]{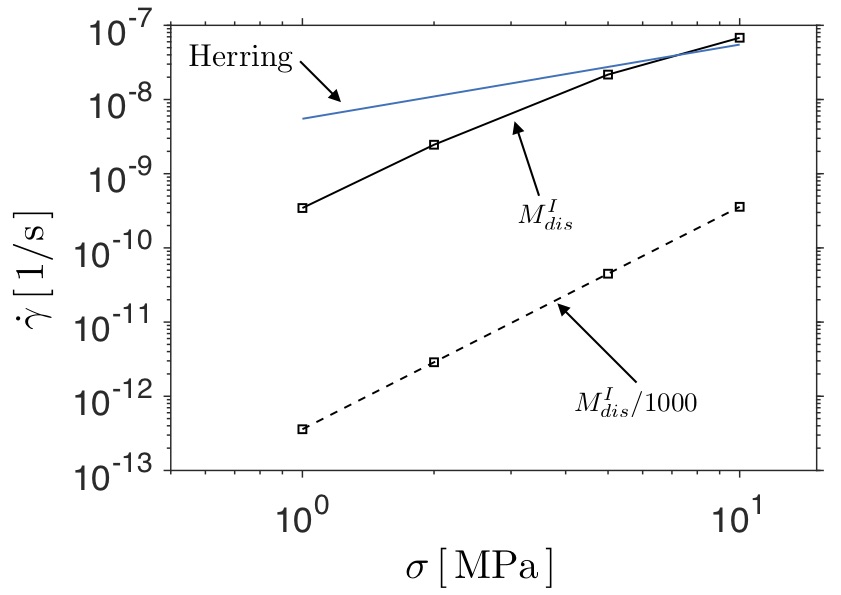}} 
		\end{tabular}
\caption{\em{ Log-log plot of steady-state creep rate against grain size (a) and against applied stress (b) at temperature $T=1100$ K. Numerical predictions with different mobilities are compared with the analytical formula by Herring Eq. (\ref{eq:Herring}). Squared marker identifies individual results obtained for the simulated grain size and applied stress.}}
\label{fig:grain_stress_bulk}
\end{figure} 

Stress dependence typical of reaction controlled creep is obtained for $M_{dis}/1000$ as well, as shown in Fig. \ref{fig:grain_stress_bulk}b. However, a similar slope is also obtained for high dislocation mobility where simulations are in disagreement with Herring's formula. Therefore, in case of diffusion through the lattice and in the range of prescribed stress and grain sizes, creep is not diffusion controlled even for high dislocation mobility.  This conclusion relies on Eq. \eqref{eq:creep_regime} which clearly shows that  the resulting creep regime does not depend exclusively on the dislocation mobility, but also on applied stress, grain size, and atomic diffusivity in GB regions.

\subsection{Temperature dependence}

The influence of applied temperature on steady-state diffusional creep is investigated in this last numerical example. Simulations are performed adopting the same geometry and boundary conditions used in Section \ref{subsec:case1}, i.e. assuming grain size $d=100$ $\mu$m and constant applied stress $\sigma=10$ MPa. 
In addition, following the discussion of the previous sections, the dislocation mobility is set equal to the intrinsic mobility $M_{dis}^I$ in order to retrieve the diffusion controlled regime. To evaluate the role played by temperature, different analyses have been performed with applied temperature ranging from $800$ to $1300$ K 
(58 \% to 96\% times the melting temperature). 

\begin{SCfigure} [] [htbp!]
\centering
 \includegraphics[height=6cm]{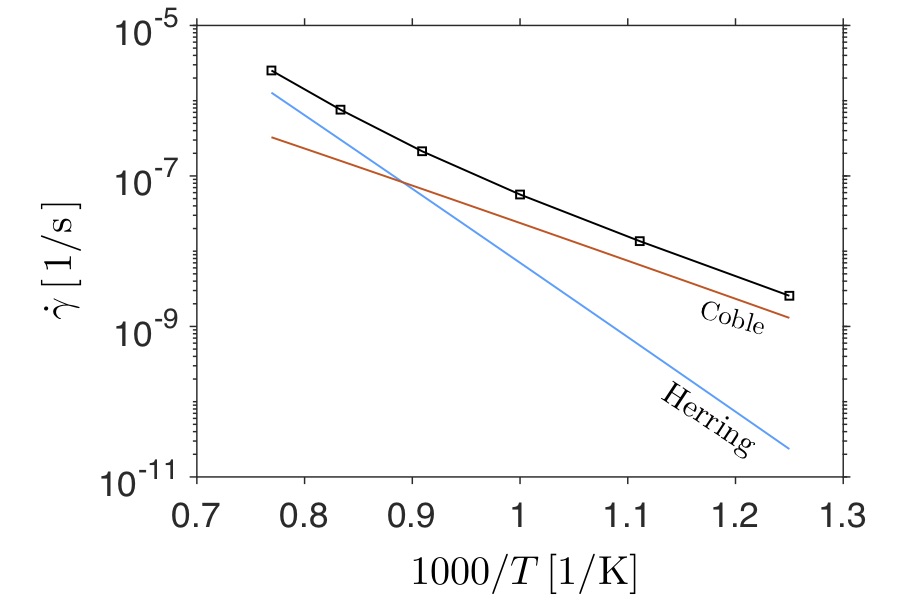}
\caption{\em{Plot of the average steady-state shear strain against temperature for a polycrystal of grain size $d=100 \, \mu$m, subjected to shear stress $\sigma =  10$ MPa,   and with assigned intrinsic mobility at GB dislocations. Numerical predictions are compared with the analytical formulas by Coble (\ref{eq:Coble}) and Herring Eq. (\ref{eq:Herring}). Squared marker identifies individual results obtained for different applied temperatures.} }
\label{fig:Tdep}
\end{SCfigure}

The simulated steady-state creep rate is plotted against the applied temperature in Figure \ref{fig:Tdep} along with the formulas by Coble \eqref{eq:Coble} and Herring \eqref{eq:Herring}.
With the selected temperature representation of the horizontal axis, the curves of Coble and Herring result in intersecting straight lines. Note that the intersection point can be viewed as the transition temperature between Coble and Herring creep. As expected, the former dominates diffusional creep for temperatures lower than the transition temperature, while Herring-like creep rules creep for higher applied temperatures.

 A similar transition from Coble to Herring creep is recovered from the numerical results as well. Indeed, for low temperatures, the simulated steady-state creep rate is in agreement with Coble's theory. Subsequently, as temperature increases, the numerical results progressively deviate from Coble's line and until tending to Herring's line. Accordingly, the transport of vacancies progressively switches from dominant boundary diffusion to diffusion through the lattice.

\subsection{Discussion}

Simulations have been performed imposing conditions such that diffusional creep is the dominant deformation mechanism, i.e. for low applied stress and high homologous temperature. In such conditions, the impact of GB dislocation mobility on creep has been studied in several case studies. 

Numerical outcomes have shown the model capability of simulating the typical grain size and stress dependence of diffusion controlled and reaction controlled creep. Owing to the general form of the employed reaction kinetics \eqref{eq:gov_climb_GBdisl}, these results are obtained naturally by tuning the mobility of GB dislocations. For the considered grain sizes and applied loads, boundaries tend to behave as perfect sources and sinks for vacancies if the intrinsic mobility is employed. Decreasing the mobility of dislocations implies a transition towards reaction controlled creep.
Moreover, it has been shown that the creep regime does not depend exclusively on the dislocation mobility. In fact, the GB reaction kinetics is also ruled by dislocation density, applied stress, grain size, and GB atomic diffusion.

The model allows the transient macroscopic response (or primary creep) to be investigated in view of the resulting kinetics of diffusional creep. It was observed that transient creep is influenced by the mobility of boundary dislocations. In particular, the transient regime is suppressed for low dislocation mobility since a steady-state condition is reached immediately after the application of the load.

In this model, grains are separated by diffuse grain boundaries. Although this may constitute a model limitation, since a realistic GB thickness cannot be assigned, such a choice allows us to study effectively the chemo-mechanical response at GB junctions. Indeed, there is no need of introducing additional constraints at boundary junctions, as usually pursued in models that implement sharp GBs. 

To counterbalance the choice of a large diffuse GB thickness, the atomic diffusivity at GBs has been reduced accordingly. This affect the GB dislocation mobility as well, in view of Eq. \eqref{eq:intrinsic_mobility}. Therefore, the modified GB diffusivity does not impact on the interplay between vacancy diffusion and operation of vacancies emission/absorption in GBs.

If dislocation mobility is sufficiently high, the evolution of vacancies and normal stress highlights a strong coupling between diffusion and mechanics. In such a condition, stress and vacancy concentration do not distribute uniformly in boundaries. Stress relaxes at GB junctions where diffusion is fast and do not limit the GB reaction kinetics. While far from boundary junctions, diffusion limits the kinetics of dislocation climb causing the stress to increase.

\section{Conclusions} \label{sec:conclusions}

A coupled diffusion-mechanical framework to study diffusional creep of polycrystalline solids has been developed and numerically implemented. The model is based on thermodynamics and dislocation physics, and relies on the assumption that diffusional creep stems from climb of dislocations along grain boundaries under applied stress \cite{Ashby1969, Ashby1972} which induces generation/annihilation of vacancies.

The governing equations of the problem have been derived in accordance with thermodynamic principles. In particular, the driving force of climb of GB dislocations has been identified as the sum of forces of mechanical and chemical nature. The kinetics of dislocation climb has been derived from physically-based mechanisms, enforcing a dependence on density and mobility of boundary dislocations. 

The classical limits of diffusion controlled and interface controlled creep have been discussed in view of the proposed governing equations. The occurrence of these two creep regimes is analyzed by comparing the kinetics of diffusion and vacancy emission/absorption in GBs. The competition between these two mechanisms induces a creep response depending on the applied conditions --- i.e. applied stress and temperature ---, polycrystal microstructure --- i.e. grain size, grain boundary network--- and crystal properties such as dislocation and vacancy mobilities, molar volume of atoms, Burgers' vector, elastic properties and activation energies.

Quantitative results of the proposed theory have been discussed in the second part of the paper where several representative numerical examples of creep of polycrystalline solids are reported. Firstly, the impact of dislocation mobility on diffusional creep has been analyzed, showing the different chemo-mechanical response at grain boundaries  and GB junctions depending on the attained creep regime. In particular, in conditions such that diffusion controlled creep is attained, vacancies accumulate/deplete at GB and GB junctions are regions of stress relaxation. Secondly, the influence of applied stress and grain size on the simulated steady-state creep rate has been highlighted. The numerical outcomes have been compared with the classical theories of Herring \cite{Herring1950} and Coble \cite{Coble1963}, obtaining good agreement when a high dislocation mobility is employed. 
Finally, the dependence of temperature is investigated, showing a continuous transition from Coble to Herring creep as temperature increases. 

The prediction of the transient evolution and steady-state distribution of stress and vacancy concentration is potentially crucial for the simulation of failure (tertiary creep) of polycrystalline aggregates. In this regard, simple phenomenological models of ductile fracture, e.g. Gurson-based model \cite{TVERGAARD1984157}, where stress and plastic-strain drive nucleation and growth of voids, could be coupled to the present formulation to predict failure. Alternatively, keeping track of vacancy concentration would allow the application of physically-based models of ductile fracture where void growth is induced by vacancy condensation \cite{CUITINO1996427}.

\section*{Acknowledgment}
The authors gratefully acknowledge the support provided by the Luxembourg National Research Fund (FNR), Reference No. 12737941.

\bibliographystyle{unsrt}


\clearpage
\begin{appendices}

\section{Numerical implementation} \label{app:num_implementation}

\subsection{Non-dimensional governing equations}

Governing equations \eqref{eq:governing} are rewritten in non-dimensional form prior to their numerical implementation. The following adimensional variables are then introduced

\begin{equation} \label{eq:nondim_var}
x_i^* = 	\frac{x_i}{\bar{l}} \, , \qquad t^* = \frac{t}{\bar{t}} \, ,  \qquad c_V^* = \frac{c_V}{\bar{c}} \, , \qquad c_L^* = \frac{c_L}{\bar{c}} \, ,\qquad u_i^* = \frac{u_i}{\bar{l}} \, , \qquad  \sigma_i^* = \frac{\sigma_i}{\bar{\sigma}} \, ,
\end{equation} 

\noindent
where $\bar{l}$, $\bar{t}$, $\bar{c}$, $\bar{\sigma}$ are reference length, time, concentration, and stress, respectively. By means of definitions \eqref{eq:nondim_var}, governing equations \eqref{eq:governing} are equivalent to

\begin{subequations} \label{eq:nondim_governing}
\begin{eqnarray}
 &&
\label{non_dim_massbalanceequation}
 \frac{ \partial c_{L}^* }{ \partial t^*} \, +  \text{div}^* \left[ \vec{h}^*_V \right] = \phi_{GB} s_V^* \, , \\ \nonumber
\\ &&
\label{non_dim_stressbalancess}
\text{div}^* \left[ \boldsymbol{\sigma}^* \right] = \vec{0} \, , \\ \nonumber
\\ &&
\label{eq:evo_beta_non}
\frac{ \partial \beta }{ \partial t^*} =  L_{GB}^* \left( t_n^* - \frac{\mu^*}{v_A^*} \right) \, .
 \end{eqnarray}
\end{subequations}

\noindent
where

\begin{equation*}
\text{div}^* \left[ \vec{h}^*_V \right] = \sum_{i=1}^3 \frac{\partial h^*_{Vi}}{ \partial x^*_i} \, ,  \qquad \text{div}^* \left[ \boldsymbol{\sigma}^* \right] = \sum_{i=1}^3 \sum_{j=1}^3 \frac{\partial \sigma^*_{ij}}{ \partial x^*_j} \vec{e}_i \, , 
\end{equation*}

\noindent
and

\begin{equation*}
\vec{h}_V^* = \frac{\bar{t} \, \vec{h}_V}{\bar{c} \, \bar{l}}  \, , \qquad s_V^* = \frac{\bar{t} \, s_V }{\bar{c}}  \, \qquad  L^*_{GB} = L_{GB} \, \bar{t} \, \bar{\sigma} \, , \qquad  t_n^* = \frac{t_n}{\bar{\sigma}} \, , \qquad \mu^* = \frac{\mu \, \bar{c} }{ \bar{\sigma}} \,, \qquad v_A^* = v_A \, \bar{c} \, .
\end{equation*}

\noindent
Equations \eqref{eq:nondim_governing} have the same form of equations \eqref{eq:governing}, but are formulated in terms of non-dimensional variables.  Similarly, the non-dimensional counterpart of the constitutive laws  \eqref{eq:const_diff_potential}, \eqref{eq:const_stress_tensor}, and \eqref{eq:const_vacancies_flux} can be easily derived by means of \eqref{eq:nondim_var}, obtaining\footnote{It is assumed that $E_V^{bulk} = E_V^{GB} = E_V$ for simplicity.}

\begin{gather*}
\mu^* = E_V^*  + \, (RT)^* \, \text{ln} \left[ \frac{c^*_V}{c^*_L }\right] - \omega^*_V \, \text{tr} \left[ \boldsymbol{\sigma}^* \right] \, , \\
\\
\boldsymbol{\sigma}^* = \mathds{C}^* :  \boldsymbol{\varepsilon}^{el}  \, ,  \\
\\
\vec{h}^*_V =  - ( 1 - \phi_{GB})   \, \Big[ \boldsymbol{D}_V^{bulk *}  \nabla^*\left[ c^*_V \right]  -   \boldsymbol{D} ^{bulk *}_\Sigma \, c^*_V  \, \nabla^* \left[ \text{tr} [ \boldsymbol{\sigma}^*] \right]  \Big]  - \phi_{GB} \, \Big[   \boldsymbol{D}_V^{GB *}  \nabla \left[ c^*_V \right]  -  \boldsymbol{D}^{GB *} _\Sigma \, c^*_V \, \nabla^* \left[ \text{tr} [ \boldsymbol{\sigma}^*] \right] \Big] \, . 
\end{gather*}

\noindent
where

\begin{subequations} \nonumber
\begin{gather}
E_V^* = \frac{E_V \, \bar{c} }{\bar{\sigma}} \, , \qquad  (RT)^* = \frac{RT \, \bar{c} }{\bar{\sigma}} \, , \qquad \omega_V^* = \omega_V \, \bar{c} \, , \qquad \mathds{C}^* = \frac{\mathds{C}}{\bar{\sigma}} \, ,  \\
\\
\boldsymbol{D}_V^{bulk *} = \frac{\boldsymbol{D}_V^{bulk } \,  \bar{t}}{\bar{l}^2} \, , \qquad \boldsymbol{D}_V^{GB *} = \frac{\boldsymbol{D}_V^{GB} \,  \bar{t}}{\bar{l}^2} \, , \qquad \boldsymbol{D} ^{bulk *}_\Sigma = \frac{\boldsymbol{D}_V^{bulk *} \omega_V \bar{\sigma} }{R T } \, , \qquad \boldsymbol{D} ^{GB *}_\Sigma = \frac{\boldsymbol{D}_V^{GB *} \omega_V \bar{\sigma} }{R T }
\end{gather}
\end{subequations}

\subsection{Weak form}
\textit{FEniCS} \cite{FenicsProject} is an open-source computing platform for solving PDEs through the Finite Element Method. The governing equations have to be formulated in variational (or weak) form and implemented in symbolic language. The weak form results from multiplying the strong form of governing equations \eqref{eq:nondim_var} by a suitable set of tests functions and performing an integration upon the domain, exploiting the \textit{integration by parts} formula with the aim of reducing the order of differentiation in space.

Note that the first order derivative of $\vec{h}_V$, in Eq. \eqref{non_dim_massbalanceequation}, can be eliminated by applying the integration by parts. However, its constitutive definition contains the second order derivative of displacement field $\vec{u}$. The latter is undetermined in standard finite element since $\vec{u}$ is approximated with global $\mathcal{C}^0$ polynomials. To include the effect of stress gradient, we then follow the approach adopted in \cite{BowerGuduruIMSM2012}, in which an additional variable $\Sigma$ is defined as 

\begin{equation} \label{eq:additional_degree}
\Sigma = \text{tr} \left[ \boldsymbol{\sigma} \right]  \, ,
\end{equation}

\noindent
which will be considered as an independent field variable from now on. Eq. \eqref{eq:additional_degree} is then added to the set of governing equations \eqref{eq:nondim_var} for the numerical implementation of the problem. 

The overall weak form of the problem is derived considering each governing equation separately at first. In what follows, the asterisk is omitted for the sake of readability. From the mass balance Eq. \eqref{non_dim_massbalanceequation}, one obtains

\begin{equation} \label{eq:weak_diff}
\begin{aligned}
  & \int_{\Omega} \hat{c}_V \left\{  \frac{ \partial c_V }{ \partial t} + \divergence{\vec{h}_V} \right\} \, \, \text{d}V = \\
 & = \,    \int_{\Omega} \hat{c}_V \,  \frac{ \partial c_V }{ \partial t}  \,  \text{d}V  +   \int_{\Omega} ( 1 - \phi_{GB})  \gradient{\hat{c}_V} \cdot \bigg\{ \boldsymbol{D}_V^{bulk}  \nabla\left[ c_V \right] - \boldsymbol{D} ^{bulk }_\Sigma \, c_V  \, \nabla\left[ \Sigma \right] \bigg\} \, \text{d}V \, +  \\
 &+   \int_{\Omega} \phi_{GB} \gradient{\hat{c}_V} \cdot \bigg\{ \boldsymbol{D}_V^{GB}  \nabla\left[ c_V \right] - \boldsymbol{D} ^{GB}_\Sigma \, c_V  \, \nabla\left[ \Sigma \right] \bigg\} \, \text{d}V \, +    \,  \int_{\partial^N \Omega_h} \hat{c}_V \, \overline{h}_V \, \text{d}A \, = 0 \, .
\end{aligned}
\end{equation}

\medskip

\noindent
The weak form of Eq. \eqref{non_dim_stressbalancess} reads

\begin{equation}  \label{eq:weak_stress} 
\begin{aligned} 
  & \int_{\Omega} \vect{\hat{u}}  \, \cdot \, \divergence{\tensor{\sigma}}  \, \text{d}V  =  -   \, \int_{\Omega} \,  \nabla_S \big[ \, \vect{\hat {u}} \,  \big]  :  \tensor{\sigma} ( c_V , \vect{u}, \Sigma , ) \,  \text{d} V \, +  \,  \int_{\partial^N \Omega_\sigma} \, \vect{\hat{u}} \,  \cdot \, \vec{ \overline{t}} \,  \text{d}A \, = 0 \,  .
\end{aligned}
\end{equation}

\medskip 

\noindent
From equation \eqref{eq:additional_degree}, one obtains

\begin{equation*} 
\int_{\Omega} \hat{\Sigma} \Big\{  \Sigma - \trace{\tensor{\sigma} ( c_V , \vect{u}, \Sigma  )} \Big\} \, \text{d}V = 0 \, .  
\end{equation*}

For the numerical implementation in \textit{FEniCS}, it was convenient to solve Eq. \eqref{eq:evo_beta_non} in weak form as well. To avoid the usage of the logarithm appearing in the definition of $\mu$, \eqref{eq:evo_beta_non} has been rewritten in the following equivalent form

\begin{equation*}
c_V - c_L  \, \text{exp}\left[ - \frac{E_V}{RT} + \frac{\omega_V \Sigma }{RT} + \frac{ v_A \, t_n( c_V , \vect{u}, \Sigma ) }{RT} - \frac{v_A \,\dot{\beta}}{L_{GB} \, RT } \,  \right] = 0 \, ,
\end{equation*}

\noindent
whose weak form is simply

\begin{equation*} 
\int_{\Omega} \hat{\beta} \left\{  c_V - c_L  \, \text{exp}\left[ - \frac{E_V}{RT} + \frac{\omega_V \Sigma }{RT} + \frac{ v_A \, t_n( c_V , \vect{u}, \Sigma ) }{RT} - \frac{\partial \beta}{\partial t}\frac{v_A \,}{L_{GB} \, RT } \,  \right] \right\} \, \text{d}V = 0 \, .  
\end{equation*}

Note that in equations \eqref{eq:weak_diff} and \eqref{eq:weak_stress}, boundary conditions \eqref{eq:NeumanBC} have been applied along with the condition that test functions $\hat{c}_V$ and $\hat{u}$, are null on the Dirichlet boundary. In conclusion the overall weak form, in the time interval $ \left[ t_0 , \,  t_f \right] $, reads

\begin{equation} 
\begin{aligned}
& \text{Find} \,  \, V = \left\{ c_V, \, \vec{u}, \, \beta , \, \Sigma \right\}   \in \mathcal{V}^{ \left[ 0 , \, t_f  \right] } \, \text{such that}  \hspace{7cm} \\
\\
  &  \int_{\Omega} \hat{c}_V \,  \frac{ \partial c_V }{ \partial t}  \,  \text{d}V  +   \int_{\Omega} ( 1 - \phi_{GB})  \gradient{\hat{c}_V} \cdot \bigg\{ \boldsymbol{D}_V^{bulk}  \nabla\left[ c_V \right] - \boldsymbol{D} ^{bulk }_\Sigma \, c_V  \, \nabla\left[ \Sigma \right] \bigg\} \, \text{d}V \, +  \\
 + &  \int_{\Omega} \phi_{GB} \gradient{\hat{c}_V} \cdot \bigg\{ \boldsymbol{D}_V^{GB}  \nabla\left[ c_V \right] - \boldsymbol{D} ^{GB}_\Sigma \, c_V  \, \nabla\left[ \Sigma \right] \bigg\} \, \text{d}V \, +  \, \int_{\Omega} \,  \nabla_S \big[ \, \vect{\hat {u}} \,  \big]  :  \tensor{\sigma} ( c_V , \vect{u}, \Sigma) \,  \text{d} V \, +  \\
+ & \int_{\Omega} \hat{\beta} \left\{  c_V - c_L  \, \text{exp}\left[ - \frac{E_V}{RT} + \frac{\omega_V \Sigma }{RT} + \frac{ v_A \, t_n( c_V , \vect{u}, \Sigma ) }{RT} - \frac{\partial \beta}{\partial t} \frac{v_A }{L_{GB} \, RT } \,  \right] \right\} \, \text{d}V \, + \\
+ & \int_{\Omega} \hat{\Sigma} \Big\{  \Sigma - \trace{\tensor{\sigma}( c_V , \vect{u}, \Sigma  )} \Big\} \, \text{d}V ,  \int_{\partial^N \Omega_h} \hat{c}_V \, \overline{h}_V \, \text{d}A  -  \,  \int_{\partial^N \Omega_\sigma} \, \vect{\hat{u}} \,  \cdot \, \vec{ \overline{t}} \,  \text{d}A  = 0 \\
\\
& \text{for all} \, \, \hat{V} = \left\{ \hat{c}_V, \, \vec{\hat{u}}, \, \hat{\beta} , \, \hat{\Sigma} \right\}   \in \mathcal{V} \, .  
\end{aligned}
\end{equation}

\noindent
 The identification of the functional space $\mathcal{V}$ falls beyond the scope of this work. 

\subsection{Discretization in time}

The evolution in time of problem \eqref{eq:weak_diff} is approximated using the \textit{Backward Euler method}. The time interval $[t_0, \, t_f] $ is divided into $N_t$ temporal steps $\Delta t = ( t_f - t_0 ) / N_t$. In addition, we define

\begin{equation} 
V |_n = V ( \vec{x}, n \Delta t ) \, , \qquad \Delta V |_{n+1} = V |_{n+1} - V |_{n}  \qquad n = 1, \, 2, \, ... \, N_t
\end{equation}

\noindent
Therefore, for any time step $n = 1, \, 2, \, ... \, N_t$, the discretized weak form in time reads

\begin{equation*} 
\begin{aligned}
& \text{Find} \, \, V |_{n+1} = \left\{ c_V  |_{n+1} , \, \vec{u}  |_{n+1}, \, \beta  |_{n+1} , \, \Sigma   |_{n+1}\right\}   \in \mathcal{V} \, \text{such that}  \hspace{7cm} \\
\\
  &  \int_{\Omega} \hat{c} \,  \frac{ \Delta c_V  |_{n+1} }{ \Delta t}  \,  \text{d}V  +   \int_{\Omega} ( 1 - \phi_{GB})  \gradient{\hat{c}_V} \cdot \bigg\{ \boldsymbol{D}_V^{bulk}  \nabla\left[ c_V |_{n+1}  \right] - \boldsymbol{D} ^{bulk }_\Sigma \, c_V |_{n+1}  \, \nabla\left[ \Sigma |_{n+1} \right] \bigg\} \, \text{d}V \, +  \\
 + &  \int_{\Omega} \phi_{GB} \gradient{\hat{c}_V} \cdot \bigg\{ \boldsymbol{D}_V^{GB}  \nabla\left[ c_V |_{n+1} \right] - \boldsymbol{D} ^{GB}_\Sigma \, c_V |_{n+1}  \, \nabla\left[ \Sigma |_{n+1} \right] \bigg\} \, \text{d}V \, + \\
 + &  \, \int_{\Omega} \,  \nabla_S \big[ \, \vect{\hat {u}} \,  \big]  :  \tensor{\sigma} ( c_V |_{n+1}, \vect{u} |_{n+1}, \Sigma |_{n+1}  ) \,  \text{d} V \, +  \\
+ & \int_{\Omega} \hat{\beta} \left\{  c_V |_{n+1} - c_L  \, \text{exp}\left[ - \frac{E_V}{RT} + \frac{\omega_V \Sigma |_{n+1} }{RT} + \frac{ v_A \, t_n( c_V |_{n+1} , \vect{u} |_{n+1}, \Sigma |_{n+1} ) }{RT} - \frac{v_A \,\Delta \beta |_{n+1}}{L_{GB} \, RT  \, \Delta t } \,  \right] \right\} \, \text{d}V \, + \\
+ & \int_{\Omega} \hat{\Sigma} \Big\{  \Sigma  |_{n+1} - \trace{\tensor{\sigma} ( c_V  |_{n+1} , \vect{u}  |_{n+1}, \Sigma  |_{n+1} )} \Big\} \, \text{d}V ,  \int_{\partial^N \Omega_h} \hat{c}_V \, \overline{h}_V  |_{n+1}  \, \text{d}A  -  \,  \int_{\partial^N \Omega_\sigma} \, \vect{\hat{u}} \,  \cdot \, \vec{ \overline{t}}  |_{n+1} \,  \text{d}A  = 0 \\
\\
& \text{for all} \, \, \hat{V} = \left\{ \hat{c}_V, \, \vec{\hat{u}}, \, \hat{\beta} , \, \hat{\Sigma} \right\}   \in \mathcal{V} \,. 
\end{aligned}
\end{equation*}

\noindent
The resulting discretized weak form has been solved in a monolithic scheme. Linear triangular elements have been selected for the spatial discretization of solution and test functions. The finite element mesh has been generated using \textit{Gmsh} \cite{gmsh2009}. To better approximate the solution fields close to GB, a finer mesh has been defined in boundary regions.

\end{appendices}

\end{document}